\documentclass[aps,prd,twocolumn,showpacs,superscriptaddress,nofootinbib]{revtex4-2}

\usepackage{graphicx}
\usepackage{dcolumn}
\usepackage{color}
\usepackage{bm}

\usepackage[english]{babel}
\usepackage{graphicx}
\usepackage{epsfig}
\usepackage{amssymb}
\usepackage{amsmath}
\usepackage[plainpages=false,pagebackref=false,pdftex]{hyperref}
\usepackage{placeins}
\hypersetup{colorlinks=true, linkcolor=blue, citecolor=blue,urlcolor=blue}
\usepackage{ulem}
\usepackage[T1]{fontenc} 
\usepackage[utf8]{inputenc}
\usepackage{color}
\usepackage[titletoc]{appendix}
\usepackage[percent]{overpic}
\usepackage{xcolor}

\begin{document}
\preprint{APS}
\title{Impact of the phase transition on  Quark-Gluon Plasma with an extremely strong magnetic field in holographic QCD}
\author{Xuanmin Cao}
\email{caoxm@jnu.edu.cn}
\affiliation{ Department of Physics and Siyuan Laboratory, Jinan University, Guangzhou 510632, China}
\author{Hui Liu}
\email{tliuhui@jnu.edu.cn}
\affiliation{ Department of Physics and Siyuan Laboratory, Jinan University, Guangzhou 510632, China}

\begin{abstract}
We investigate the phase transition within an extremely strong magnetic background field, employing a holographic Quantum Chromodynamics (QCD) model with a focus on the entropy, pressure and Polyakov loop properties. At relatively modest magnetic field strengths, our study discerns a crossover transition between the normal phase and the Quark-Gluon Plasma (QGP) phase as the temperature rises. In contrast, under the influence of an extremely strong magnetic field, a first-order phase transition is observed. A critical point is identified at $ (eB_c, T_c) \approx (2.8623 \, \text{GeV}^2, 0.1191 \, \text{GeV}) $, which corresponds to a second-order phase transition. This phase structure is found to be in qualitative agreement with lattice simulation predictions reported in [Phys. Rev. D \textbf{105}, 034511 (2022)].
Furthermore, we explore the impact of the magnetic field on the jet quenching parameter across various phases. At zero magnetic field ($ eB = 0$ ), the normalized jet quenching parameter $ \hat{q} / T^3 $ exhibits a monotonic increase with temperature. However, in the presence of a magnetic background field, the normalized jet quenching parameters not only display directional anisotropy but also experience a universal enhancement, particularly in the vicinity of the critical temperature region. This observation suggests that the jet quenching parameter could potentially act as an indicator of phase transitions.
\end{abstract}

\maketitle

\section{Introduction}
The investigation of the quantum chromodynamics (QCD) properties under a magnetic field is one of the most important fields in high energy physics ~\cite{suganuma1991manifestation,suganuma1991behavior,suganuma1993chiral,Kharzeev:2013jha,Andersen:2014xxa,Miransky:2015ava}. The background magnetic field could be generated and widely exist in non-central heavy ion collisions~\cite{Skokov:2009qp,Deng:2012pc,Jiang:2024bez}, in the early cosmology~\cite{Grasso:2000wj}, and in magnetar~\cite{Turolla:2015mwa}. Many important and interesting phenomena are found in this area, such as  the  Nielsen-Olesen instability~\cite{Nielsen:1978rm}, chiral magnetic effects~\cite{Fukushima:2008xe,Kharzeev:2007jp}, chiral magnetic wave~\cite{Kharzeev:2010gd,Wang:2021nvh,Wu:2020wem} and the inverse magnetic catalyses~\cite{Bruckmann:2013oba}.

The chiral and confining properties of QCD are directly influenced by the presence of a magnetic field. An extremely strong magnetic field can even change the order of the phase transition, which could be important for our understanding of the early universe~\cite{Witten:1984rs,Vachaspati:1991nm}. One of the most powerful methods for studying QCD in a finite magnetic background field is the lattice simulation.  Lattice simulations~\cite{DElia:2010abb,Ilgenfritz:2012fw,Ding:2020inp}, which employed standard staggered fermions, demonstrated a rise of transition temperature in chiral condensates with Polyakov loops. However, these findings are in contrast with the results of lattice simulations ~\cite{Bali:2011qj,Bornyakov:2013eya,DElia:2018xwo,Endrodi:2015oba,Endrodi:2019zrl} that incorporated improved staggered fermions.
 The discrepancy is primarily attributed to lattice cutoff effects~\cite{Ding:2020inp}. Moreover, lattice simulations have demonstrated that a first-order transition exists when $eB \gtrsim 11 m_\pi^2$ with $N_f=3$~\cite{Ding:2020inp}. Extrapolation of the lattice results predicts a critical point beyond $eB=3.25 \ {\rm GeV^2}$ with $N_f=1+1+1$~\cite{Endrodi:2015oba}. In more recent investigations, a lattice simulation for $N_f = 2 + 1$ \cite{DElia:2021yvk}, which employs stout-improved staggered fermions and try to control the cut-off effect, has indicated that the critical temperature decreases as the magnetic field strengthens. Despite the absence of technical issues such as the sign problem at finite chemical density, lattice simulations remain constrained by limitations in computational power. They observed a crossover transition at $eB=4\ {\rm GeV^2}$ and a first order phase transition at $eB=9\ {\rm GeV^2}$, indicating the presence of a critical end point between these values. 

The phase structure of QCD under magnetic fields has also been extensively investigated using low energy effective theories and models, such as the bag model~\cite{Fraga:2012fs}, chiral perturbation theory ($\chi$PT)~\cite{Andersen:2012dz,Andersen:2012zc}, the Nambu-Jona- Lasinio (NJL) model~\cite{Gatto:2010qs,Gatto:2010pt,Sheng:2021evj,Cao:2021rwx,Endrodi:2019whh} and the quark-meson (QM) model~\cite{Kamikado:2013pya}. However, a unified consensus on the phase diagram at finite magnetic fields and temperatures has not yet been reached across these various studies. The phase diagram in an extremely strong magnetic field remains an open topic. Therefore, it is necessary to conduct further research in this domain with a variety of numerical and theoretical methods.

In addition to the traditional methods, holographic duality is another powerful approach for studying strongly coupled systems~\cite{Maldacena:1997re,Gubser:1998bc,Witten:1998qj}. Given that the quark-gluon-plasma (QGP)  exists as a strongly coupled state~\cite{Shuryak:2008eq}, holographic duality has become an important tool in the QCD investigations~\cite{Kovtun:2004de,Erlich:2005qh,Sakai:2004cn,Sakai:2005yt,Karch:2006pv,Gursoy:2007cb,Gursoy:2007er,Gubser:2008yx,Gubser:2008ny,Li:2011hp,Kim:2012ey,Cai:2012xh,Li:2012ay,Li:2013oda,Chen:2020ath,Chen:2021wzj,Chen:2022goa,Liu:2023hoq,Rougemont:2023gfz,Gubser:2008ny,Gubser:2008yx,DeWolfe:2010he,Gursoy:2007cb,Gursoy:2007er}. Regarding investigations involving magnetic backgrounds, various studies have employed different  constructive models, including bottom-up holographic models  in the Veneziano limit~\cite{Gursoy:2017wzz,Gursoy:2021efc}  and Magnetized Einstein-Maxwell-Dilaton (EMD) model~\cite{Rougemont:2015oea,Finazzo:2016mhm,Dudal:2016joz,Li:2016gfn,Bohra:2019ebj,Arefeva:2020bjk,Shahkarami:2021gzl,Wen:2024hgu}. An abundance of research has been dedicated to studying the QCD phase diagram within the finite temperature and chemical potential plane \cite{DeWolfe:2010he,Critelli:2017oub,Zhao:2023gur,Cai:2024eqa,Chen:2024ckb}. However, in this study, we employ the magnetized EMD model to explore the phase transition in the context of an extended magnetic field region, especially the extremely strong magnetic field. 
However, it should be noted that the EMD model does not include hadronic degrees of freedom such as pion or rho mesons. Extending the EMD model to include flavor dynamics and calculate hadron masses remains an open question~\cite{Liu:2023pbt}. 
In this paper, we would focus exclusively on the confinement/deconfinement phase transition and the associated thermodynamic quantities under extremely strong magnetic fields. In the previous investigations, they utilized the magnetized EMD model \cite{Rougemont:2015oea,Finazzo:2016mhm} and have yielded some findings. They have predominantly concentrated on crossover transitions and have not yet provided an exhaustive examination of the domain characterized by extremely strong magnetic fields. Building upon this foundation, we have developed numerical code to extend the parameter space  of the magnetic field and temperature.
Then, we study the extremely strong magnetic field region and observe first-order phase transitions.

Taking into account the first-order transitions, we discuss their influence on the properties of QGP.  Among various QGP signals in heavy-ion collisions, jet quenching is a crucial one~\cite{Cao:2020wlm,Cao:2024pxc,Tuchin:2010vs}. The holographic calculation of the jet quenching parameter $\hat{q}$ has been previously provided ~\cite{Liu:2006ug} and extensively tested in some holographic models~\cite{Ficnar:2011yj,Rougemont:2015wca,Li:2016bbh,Rougemont:2020had,Li:2014hja,Li:2014dsa}. In this work, we focus on the behaviors of anisotropic jet quenching parameters under both crossover and first order phase transitions.

This paper is organized as follows. In section II, we introduce the construction of magnetized EMD model and  how it realizes duality to QCD. In section III, we numerically obtain the mapping between the input parameters $(\phi,\mathcal{B})$ and the physical parameters $(T,eB)$. We then study the entropy variation with temperature at fixed magnetic field, determine the phase boundary of the first-order transition and identify the critical end point in $T-eB$ plane. In addition, we calculate the Polyakov loop to do the multi-verify for the phase boundary. In section IV, we investigate the effect of the first-order transition to the jet quenching parameters. Section V is a brief conclusion.

\section{The Einstein-Maxwell-Dilation model}
In this section, we provide a brief introduction to the EMD model with a finite magnetic field $B\neq 0$ and zero chemical potential $\mu_B=0$, as constructed in Refs.~\cite{Rougemont:2015oea,Finazzo:2016mhm}.  The chiral symmetry is not shown as an explicit degree of freedom. According to the analysis in Ref.~\cite{Son:2004iv}, it is reasonable to conclude that only a particular combination of the condensate and baryon density remains as a hydrodynamic mode, making the baryon density a suitable variable to describe the system. This bottom-up approach aims to mimic the QCD system by fitting lattice results for thermodynamic of the confinement/deconfinement phase transition from the holographic perspective, a method originally proposed by Gubser and Nellore in Ref.~\cite{Gubser:2008ny}.

In the EMD model, a scalar field $\phi$, i.e., a dilaton field, and an abelian gauge field $A_\mu$ are included.  The five dimension action is
\begin{eqnarray}
   S&=&\frac{1}{2 \kappa^2}\int d^5 x\sqrt{-g}\left[R-\frac{1}{2}(\partial_\mu \phi)^2\right .\nonumber\\
   & &\left .-V(\phi)-\frac{f(\phi)}{4}F_{\mu\nu}^2\right], 
\end{eqnarray}\label{action}
with
\begin{eqnarray}
V(\phi)&=&-v_0 {\rm cosh}(v'_0 \phi)+ v_2 \phi^2+v_4 \phi^4+v_6 \phi^6,\nonumber\\
    \kappa^2&=&8\pi G_5,\nonumber\\
    f(\phi)&=&c_3 {\rm sech}(c_0+c_1 \phi+c_2\phi^2),
\end{eqnarray}
and the field strength tensor $F_{\mu\nu}=\partial_\mu A_\nu-\partial_\nu A_\mu$. The dilaton potential $V(\phi)$  and gravitational constant $\kappa^2\equiv 8\pi G_5$ is determined by fitting the lattice data of equation of state at $B=0$ in Ref.~\cite{Borsanyi:2013bia}~\footnote{For the fitting of dilaton potential, the Breitenlohner-Freedman bound should be satisfied~\cite{Breitenlohner:1982jf,Breitenlohner:1982bm}.}. The coupling between the gauge field and dilaton $f(\phi)$ is determined by fitting the lattice results of the magnetic susceptibility at $B=0$ in Ref.~\cite{Bonati:2013vba}. The values of these parameters are summarised in Table~\ref{tab:modelparameters}. For the asymptotically $\rm AdS_5$ geometries, the ultraviolet asymptotically expansion should satisfy $V(\phi\rightarrow 0)\approx -12/L^2+m^2\phi^2/2$ with $L$ the asymptotic $\text{AdS}_5$ radius, in which the mass $m$ and the scaling dimension $\Delta$ of the dual operator in the gauge theory satisfy the relation
\begin{eqnarray}
   m^2 L^2=-\nu \Delta=(4-\Delta)\Delta. 
\end{eqnarray}
Here,  the radius is always set to be unity, i.e., $L=1$. Thus, one can derive $\Delta= 2.73294$.


\begin{table}[h!]
\centering
\begin{tabular}{l|c|c|c|c|c|c|c|c|c|c}
\hline
 Params& $G_5$ & $v_0$ & $v'_0$ & $v_2$ & $v_4$ & $v_6$ & $c_0$ & $c_1$ & $c_2$ & $c_3$\\\hline
Values & $\frac{46}{100}$ & $-12$ & $\frac{63}{100}$ & $\frac{65}{100}$ & $-\frac{5}{100}$ & $\frac{3}{1000} $ & $-\frac{32}{100} $ & $-\frac{15}{100} $ & $\frac{22}{100}$ & $\frac{95}{100}$\\
\hline
\end{tabular}
\caption{\label{tab:modelparameters}An table of the parameters of the EMD model~\cite{Finazzo:2016mhm}.}
\end{table}

In this EMD model, the magnetic field $\mathcal{B}$ is introduced through the abelian gauge field at the boundary and set to be a constant in the $z$ direction. Therefore, one can take the metric ansatz as
\begin{eqnarray}\label{eq:metric}
    ds^2&=&e^{2a(r)}[-h(r)dt^2+dz^2]+e^{2c(r)}(dx^2+dy^2)\nonumber\\
    & &+\frac{e^{2b(r)}dr^2}{h(r)},\nonumber\\
    \phi&=&\phi(r),\nonumber\\
    A&=&A_\mu dx^{\mu}=\mathcal{B} xdy\Rightarrow F=dA= \mathcal{B}dx\wedge dy,
\end{eqnarray}
with five dimension coordinates $(t,x,y,z,r)$.  As a result of the magnetic field, the rotation symmetry is broken from $SO(3)$ to $SO(2)$. Along the holographic direction, the boundary of the asymptotically $AdS_5$ is at $r\rightarrow \infty$ and the horizon of the black hole at $r_H$, which defined by $h(r_H)=0$. Under this metric in~\eqref{eq:metric}, the equations of motion (EOMs) can be derived as,
\begin{widetext}
\begin{subequations}\label{eq:EOMs}
\begin{eqnarray}
    \phi''+\left (2 a'+2 c'-b'+\frac{h'}{h}\right )\phi'-\frac{e^{2b}}{h}\left(\frac{\partial V(\phi)}{\partial \phi} +\frac{\mathcal{B}^2 e^{-4c}}{2}\frac{\partial f(\phi)}{\partial \phi}\right)&=&0,\\
    a''+\left(\frac{14}{3}c'-b'+\frac{4h'}{3h}\right)a'+\frac{8}{3}a'^2+\frac{2}{3}c'^2+\frac{2h'}{3h}c'+\frac{2e^{2b}}{3h}V(\phi)-\frac{1}{6}\phi'^2&=&0,\\
    c''-\left(\frac{10}{3}a'+b'+\frac{h'}{3h}\right )c'+\frac{2}{3}c'^2-\frac{4}{3}a'^2-\frac{2h'}{3h}a'-\frac{e^{2b}}{3h}V(\phi)+\frac{1}{3}\phi'^2&=&0\\
    h''+(2a'+2c'-b')h'&=&0,
\end{eqnarray}
with a constraint equation,
\begin{eqnarray}
    a'^2+c'^2-\frac{1}{4}\phi'^2+\left(\frac{a'}{2}+c'\right)\frac{h'}{h}+4a'c'+\frac{e^{2b}}{2h}\left(V(\phi)+\frac{\mathcal{B}^2 e^{-4c}}{2}f(\phi)\right)&=&0,
\end{eqnarray}
\end{subequations}
\end{widetext}
where the prime denotes the derivative with respect to the fifth dimension $r$.

To investigate the properties of the EMD model, one should solve the EOMs in Eqs.~\eqref{eq:EOMs} and extract the thermaldynamic quantities. The first step involves obtaining the near-boundary and near-horizon asymptotic expansions for the bulk field $\phi(r)$ and $a(r), c(r), h(r)$. We adopt the domain-wall gauge $b(r)=0$. Consequently, the ultraviolet asymptotic expansions can be derived as follows:
\begin{subequations}\label{eq:UVexpansionn}
    \begin{eqnarray}
        a(r)&=&\alpha(r)+\cdots,\\
        c(r)&=&\alpha(r)+c_0^{\rm far}-a_0^{\rm far}+\cdots,\\
        h(r)&=&h_0^{\rm  far}+h_4^{\rm  far} e^{-4\alpha(r)}+\cdots,\\
        \phi(r)&=&\phi_Ae^{-\nu \alpha(r)}+\phi_B e^{-\Delta\alpha(r)}+\cdots,
    \end{eqnarray}
with 
\begin{eqnarray}
\alpha(r)&=&a_0^{\rm 
 far}+r/\sqrt{h_0^{\rm 
 far}}.
\end{eqnarray}
\end{subequations}
and $\cdots$ denote higher-order terms. The integration constants $a_0^{\rm far}$, $c_0^{\rm far}$, $h_0^{\rm far}$ and $\phi_A$ are related to the thermodynamic quantities. Near the horizon $r_H=0$, the asymptotic expansions of the bulk fields are given by:
\begin{eqnarray}\label{eq:IRexpansionn}
    Y(r)=\sum_{n=0}^{\infty}Y_n (r-r_H)^n,
\end{eqnarray}
where $Y={a,c,h,\phi}$. In the expansion coefficients, due to the freedom in rescaling properties, one can set $r_H=a_0=c_0=0$ and $h_1=1$. One can also assume $h_0=0$. After these settings and assumptions, only $\phi_0$ and $\mathcal{B}$ remain undetermined among the seven parameters in the near horizon expansions.  Once $\phi_0$ and $\mathcal{B}$ are given, the values $X(\epsilon)$~\footnote{Due to the singularity at horizon, a small deviation $\epsilon$ is used in the numerical strategy. We choose $\epsilon=10^{-15}$ in this work.} and their first-order derivative values $X'(\epsilon)$ of the near horizon expansion are determined. One can numerically solve the EOMs in Eqs~\eqref{eq:EOMs}  by fitting the UV asymptotic expansions in Eq.~\eqref{eq:UVexpansionn} with these numerical solutions to obtain the corresponding integration constants $a_0^{\rm far},c_0^{\rm far}, h_0^{\rm far}$ and $\phi_A$. From our numerical tests, we find that choosing the maximum radial coordinate $r_{max}=10$ is sufficient for the near-boundary asymptotic expansions. Thus, the numerical solutions are all given in the interval $r\in [\epsilon, r_{max}]$.

As previously shown, we can obtain the numerical results with $\phi_A\neq 0$, which can be understood as renormalization group flows induced  by a slight deformation of a relevant operator. However, to accurately mimic QCD, $\phi_A$ must be maintained constant.  To accomplish this, it is necessary to perform a coordinate transformation from the numerical coordinates to the standard coordinates. In the standard coordinates, the metric and the UV asymptotic expansions are
\begin{eqnarray}
 d\tilde{s}^2&=&e^{2\tilde{a}(\tilde{r})}[-\tilde{h}(\tilde{r})d\tilde{t}^2+d\tilde{z}^2]+e^{2\tilde{c}(\tilde{r})}(d\tilde{x}^2+d\tilde{y}^2)+\frac{d\tilde{r}^2}{\tilde{h}(\tilde{r})},\nonumber\\
    \tilde{\phi}&=&\tilde{\phi}(\tilde{r}),\nonumber\\
    \tilde{A}&=&\tilde{A}_\mu d\tilde{x}^\mu=\hat{B}\tilde{x}d\tilde{y}\Rightarrow \tilde{F}=d\tilde{A}=\hat{B}d\tilde{x}\wedge d\tilde{y},\nonumber\\
\end{eqnarray}
and 
\begin{subequations}\label{eq:UVexpansions}
\begin{eqnarray}
    \tilde{a}(\tilde{r})&=&\tilde{r}+\cdots,\\
    \tilde{c}(\tilde{r})&=&\tilde{r}+\cdots,\\
    \tilde{h}(\tilde{r})&=&1+\cdots,\\
    \tilde{\phi}(\tilde{r})&=&e^{-\nu \tilde{r}}+\cdots.
\end{eqnarray}
\end{subequations}
with the boundary at $\tilde{r}\rightarrow \infty$ and the horizon at $\tilde{r}=\tilde{r}_H$. The numerical coordinates and the standard coordinates are equivalent and satisfy
\begin{equation}\label{eq:relation}
    d\tilde{s}^2=ds^2,\quad\quad \tilde{\phi}(\tilde{r})=\phi(r),\quad\quad \hat{B}d\tilde{x}\wedge d\tilde{y}=\mathcal{B} dx\wedge dy.
\end{equation}
Using the relation in Eq.~\eqref{eq:relation} and comparing the UV asymptotic expansions in numerical and standard coordinates in Eqs.~\eqref{eq:UVexpansionn} and ~\eqref{eq:UVexpansions}, one can derive
\begin{subequations}
\begin{eqnarray}
   &\tilde{r}=\frac{r}{\sqrt{h_0^{\rm far}}}+a_0^{\rm far}-\ln(\phi_A^{1/\nu}),\\
&\tilde{t}=\phi_A^{1/\nu}\sqrt{h_0^{\rm far}}t,\\
   &\tilde{x}=\phi_A^{1/\nu} e^{c_0^{\rm far}-a_0^{\rm far}}x,\\
  &\tilde{y}=\phi_A^{1/\nu} e^{c_0^{\rm far}-a_0^{\rm far}}x,\\
    &\tilde{z}=\phi_A^{1/\nu}z,
\end{eqnarray}    
\end{subequations}
\begin{subequations}\label{eq:explicitrelation}
    \begin{eqnarray}
       & \tilde{a}(\tilde{r})=a(r)-\ln (\phi_A^{1/\nu}),\\
       & \tilde{c}(\tilde{r})=c(r)-(c_0^{\rm far}-a_0^{\rm far})-\ln (\phi_A^{1/\nu}),\\
       &\tilde{h}(\tilde{r})=\frac{h(r)}{h_0^{\rm far}},\\
       &\tilde{\phi}(\tilde{r})=\phi(r),
    \end{eqnarray}
\end{subequations}
and
\begin{equation}\label{eq:explicitrelationb}
    \hat{B}=\frac{e^{2(a_0^{\rm far}-c_0^{\rm far})}}{\phi_A^{2/\nu}}\mathcal{B}.
\end{equation}

\begin{figure}
\centering
\includegraphics[width=0.9\linewidth]{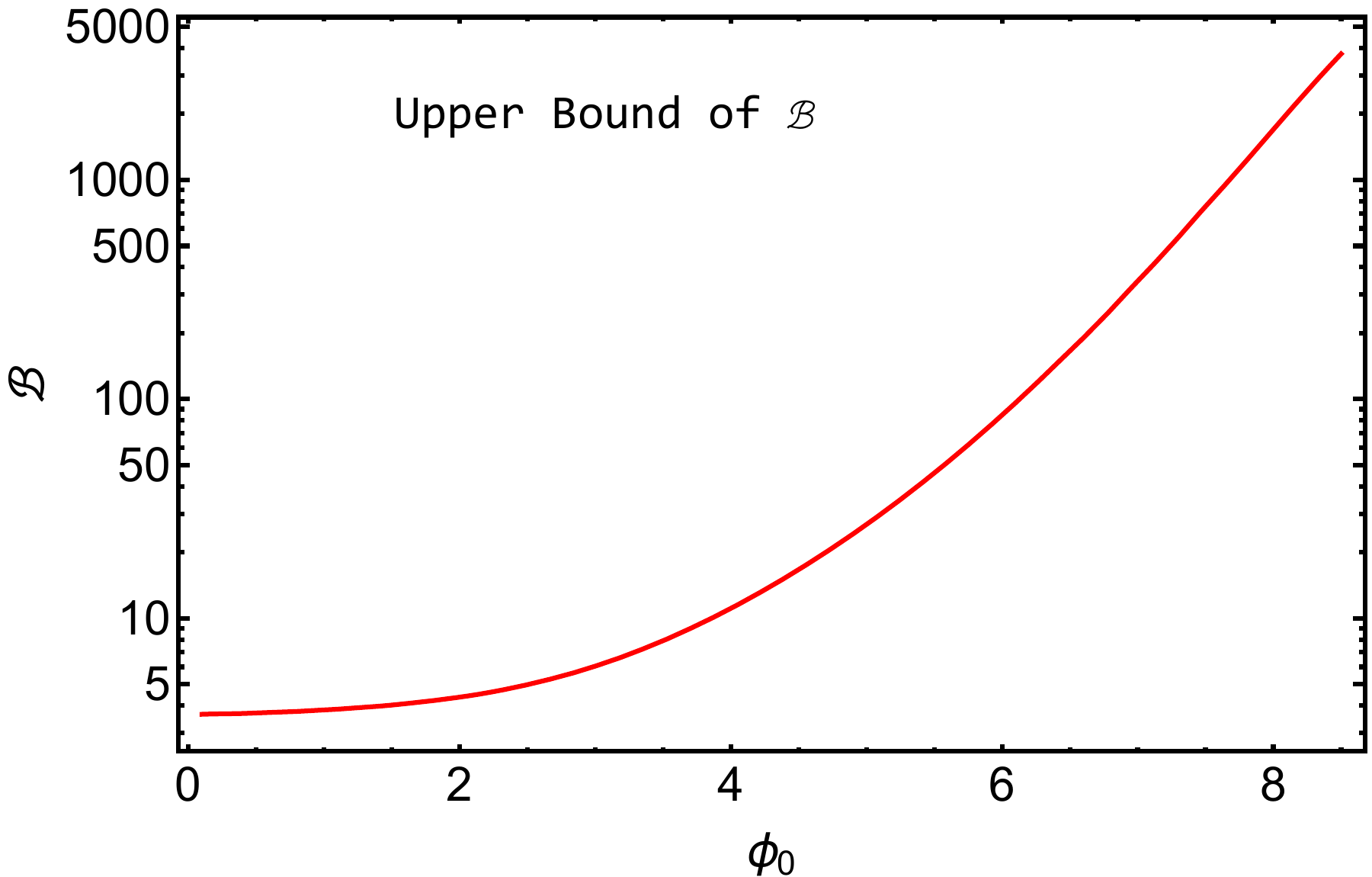}
\includegraphics[width=0.9\linewidth]{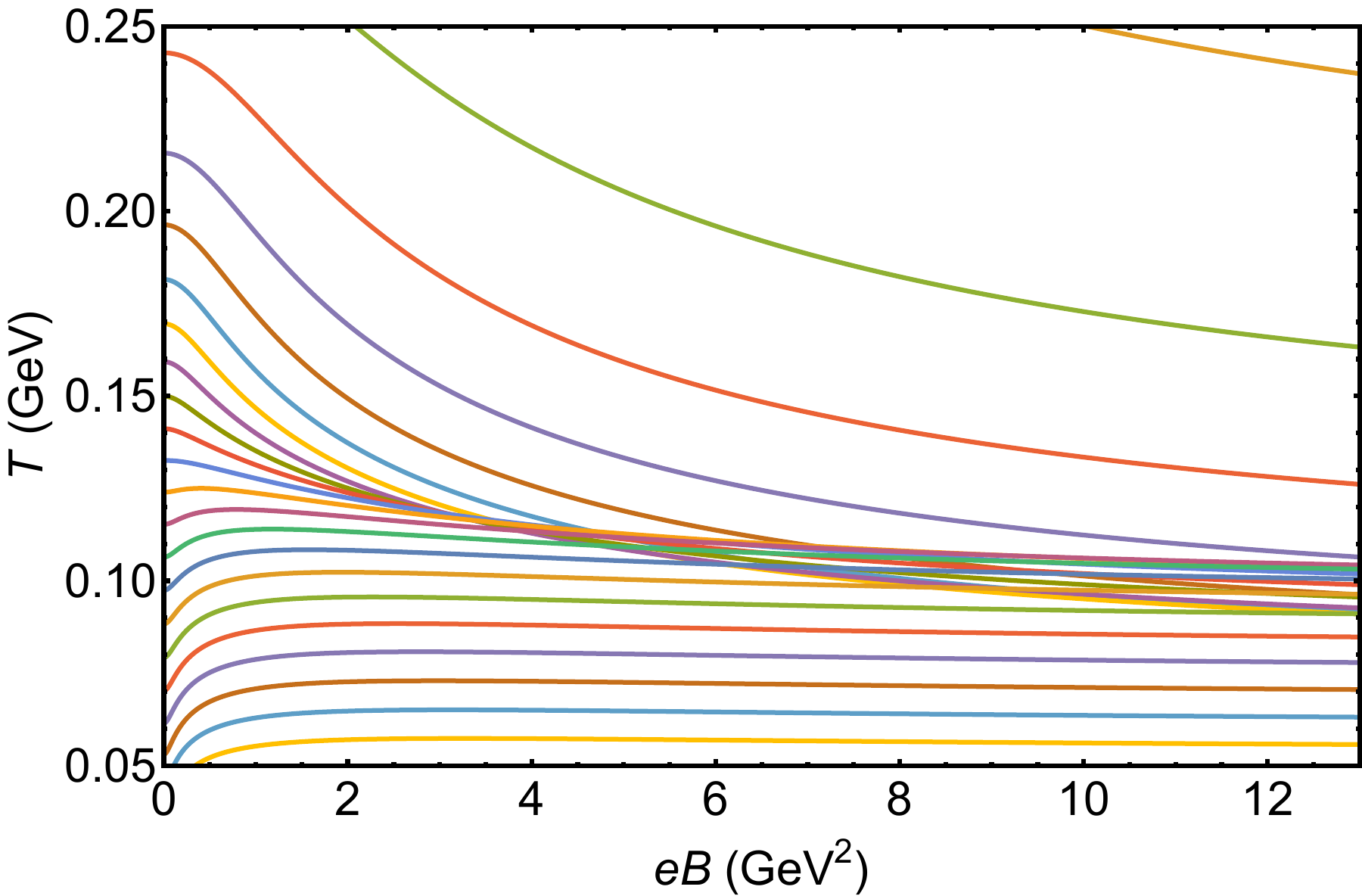}
\caption{\label{fig:phibtotb} The upper panel is the  upper bound of the input parameter $\mathcal{B}$ at a given $\phi_0$. In the lower panel, each curve of $(T,eB)$ corresponding to a fixed $\phi_0$.}
\end{figure}

Therefore, one can solve the EOMs in the  numerical coordinates and obtain the numerical solutions  of the QCD quantities in the standard coordinates through the relations in Eqs.~\eqref{eq:explicitrelation} and~\eqref{eq:explicitrelationb}. In the upper panel of Fig.~\ref{fig:phibtotb}, the curve represents the upper bound of the the input parameters $(\phi_0,\mathcal{B})$.
From the relation in Eqs.~\eqref{eq:explicitrelation} and asymptotic expansions at horizon in Eqs.~\eqref{eq:IRexpansionn}, one can obtain the QCD temperature as
\begin{equation}
   \hat{T}=\left.\frac{\sqrt{-\tilde{g}_{\tilde{t}\tilde{t}}^{'}\tilde{g}^{'\tilde{r}\tilde{r}}} }{4\pi}\right |_{\tilde{r}=\tilde{r}_H}=\frac{e^{\tilde{a}(\tilde{r}_H)}}{4\pi}|\tilde{h}'(\tilde{r}_H)|=\frac{1}{4\pi \phi_A^{1/\nu}\sqrt{h_0^{\rm far}}},
\end{equation}

\begin{figure}
    \centering
    \includegraphics[width=0.9\linewidth]{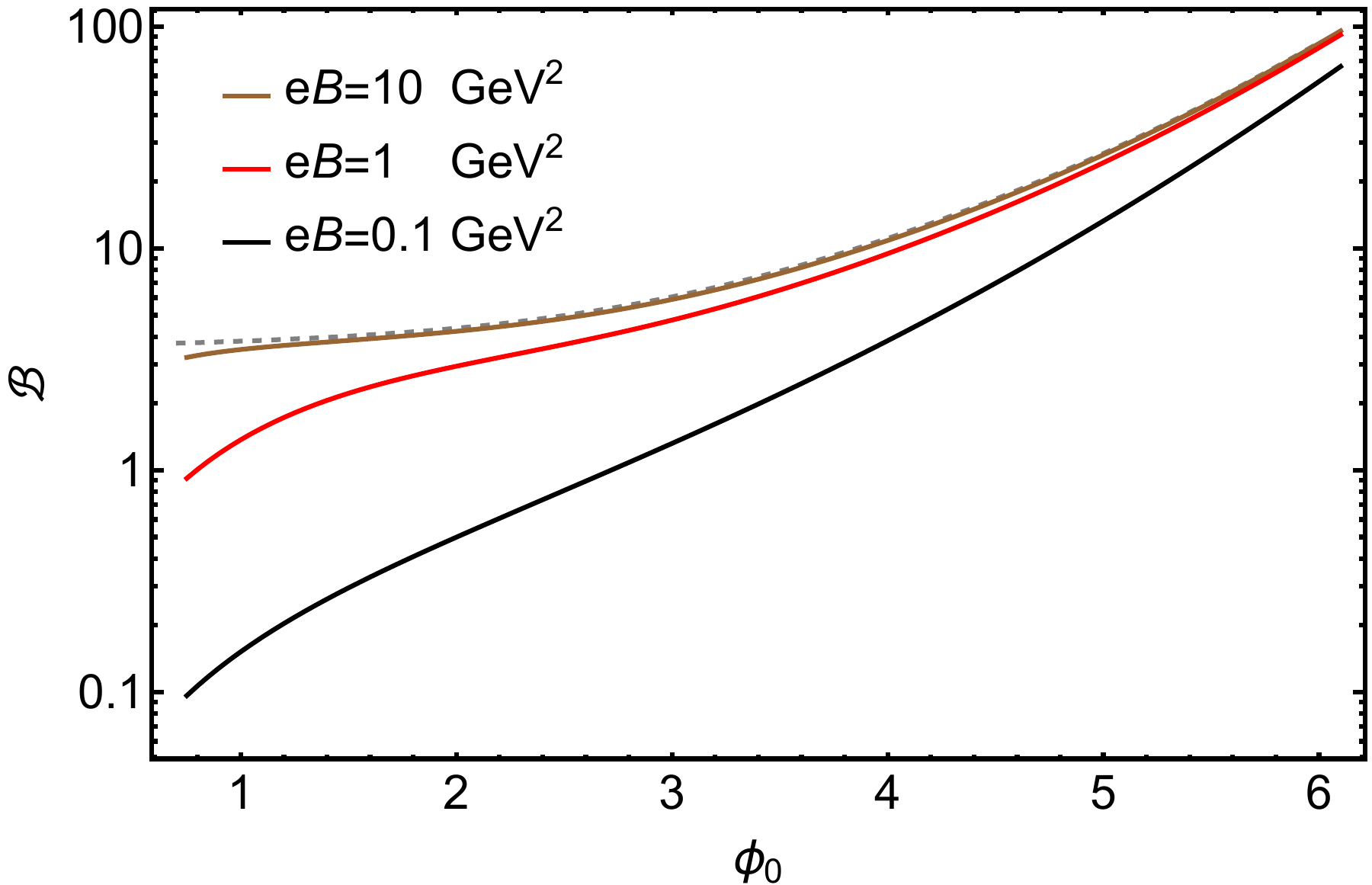}
    \caption{The input parameters of ($\phi_0,\mathcal{B}$) under the condition of fixed magnetic field $B$.}
    \label{fig:mappingbackwithfixedB}
\end{figure}

To mimic to the real QCD, the temperature corresponding to the minimum of the sound speed $c_s^2$ in the holographic results is compared to the lattice results. Based on this comparison, a scaling factor is determined,
\begin{equation}
    \lambda=\frac{T^{\rm lattice}_{\rm min\ c_s^2}}{\hat{T}_{\rm min\ c_s^2}}\approx 1058.83\quad {\rm MeV}.
\end{equation}
Therefore, the QCD quantities $X$ can be obtained by scaling $\hat{X}$ with $\lambda$, following the relation $X=\hat{X}\lambda^p\ [{\rm{MeV}}^p]$, where $p$ is determined by dimensional analysis. For example $T=\hat{T}\lambda\ [{\rm MeV}]$ and $B=\hat{B}\lambda^2\ [{\rm MeV}^2] $.
In the lower panel of Fig.~\ref{fig:phibtotb}, we show the numerical results for ($T, eB$), where each curve corresponds to a fixed $\phi_0$ and a series of $\mathcal{B}$. Normally, small $\phi_0$ maps to high $T$, large $\phi_0$ maps to low $T$ and small $\mathcal{B}$ maps to small $eB$. As $\mathcal{B}$ approaches its upper bound, a slight variation in $\mathcal{B}$ induces a large change of $eB$. Noticing that the fixed $\phi_0$ curves in  lower panel of Fig.~\ref{fig:phibtotb} are convex for large $\phi_0$ ( i.e., low $T$) but concave for small $\phi_0$ ( i.e., high $T$), these curves may exhibit overlaps at the moderate temperatures in the range of $T=100\sim 150$ MeV.
\begin{figure}
    \centering    
    \includegraphics[width=0.85\linewidth]{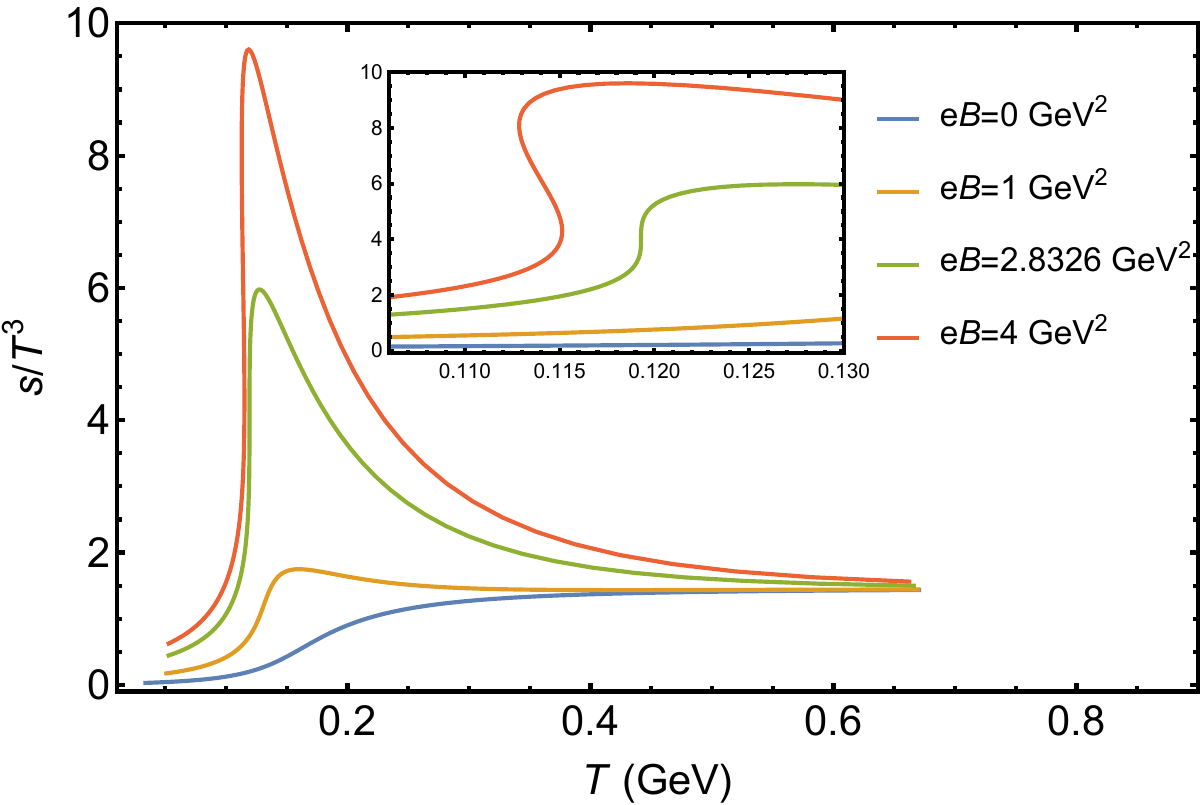}
    \includegraphics[width=0.97\linewidth]{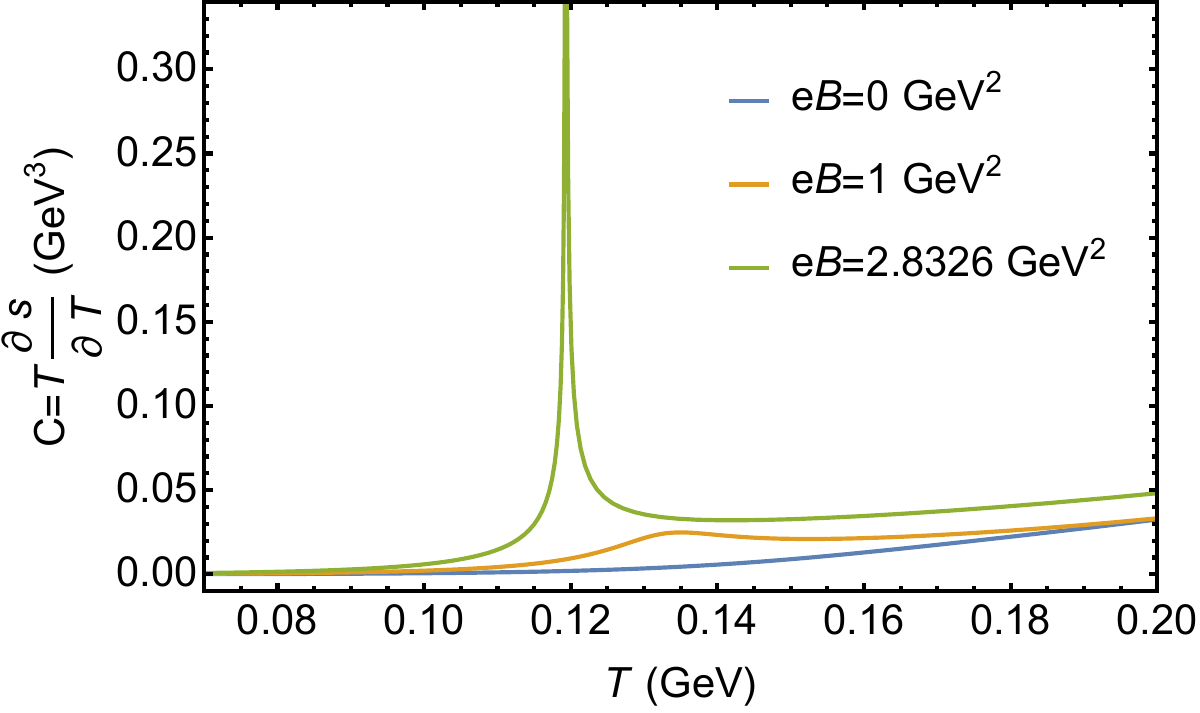}
\caption{\label{fig:entropy} The upper panel is the normalized entropy density $s/T^3$ as a function of temperature with different fixed magnetic fields. The inset is a enlarge in the region $T\in [0.106,0.123]$  GeV. The lower panel is the specific heat with different fixed magnetic fields.}
\end{figure}
\begin{figure*}
    \centering
    \includegraphics[width=0.42\linewidth]{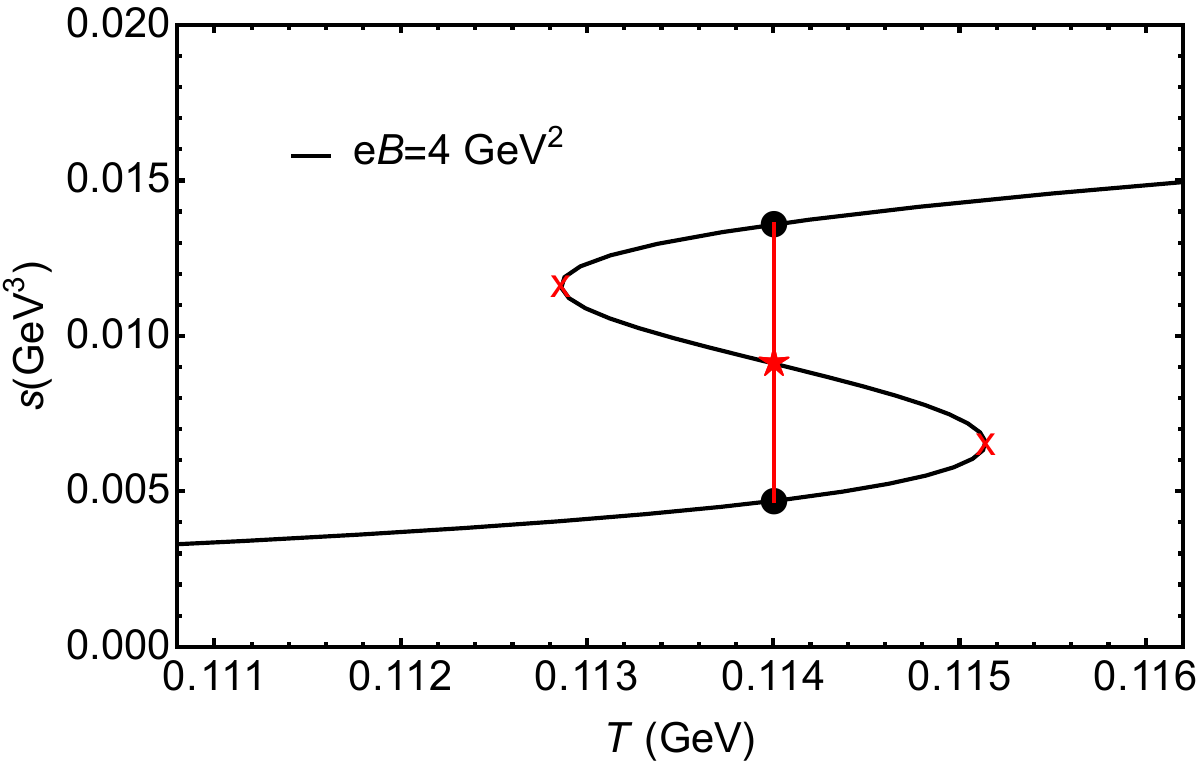}
    \hspace{0.5cm}
    \includegraphics[width=0.405\linewidth]{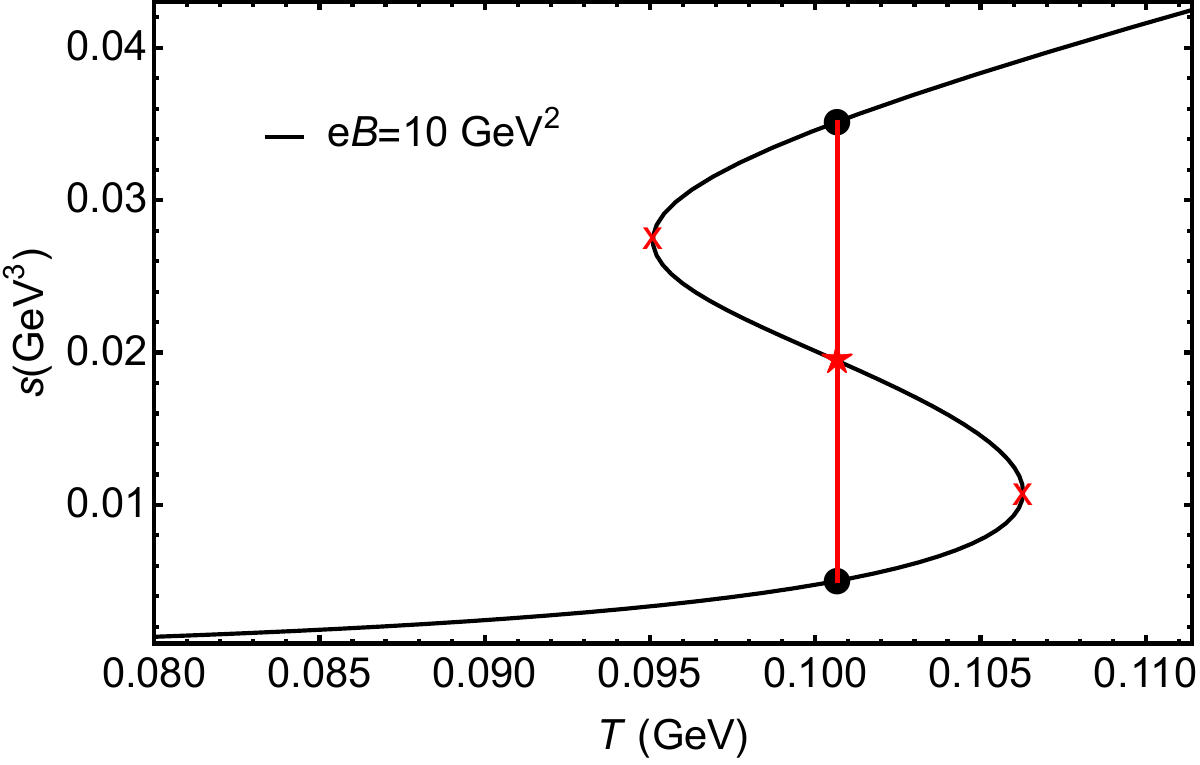}
    \hspace{0.2cm}  \includegraphics[width=0.435\linewidth]{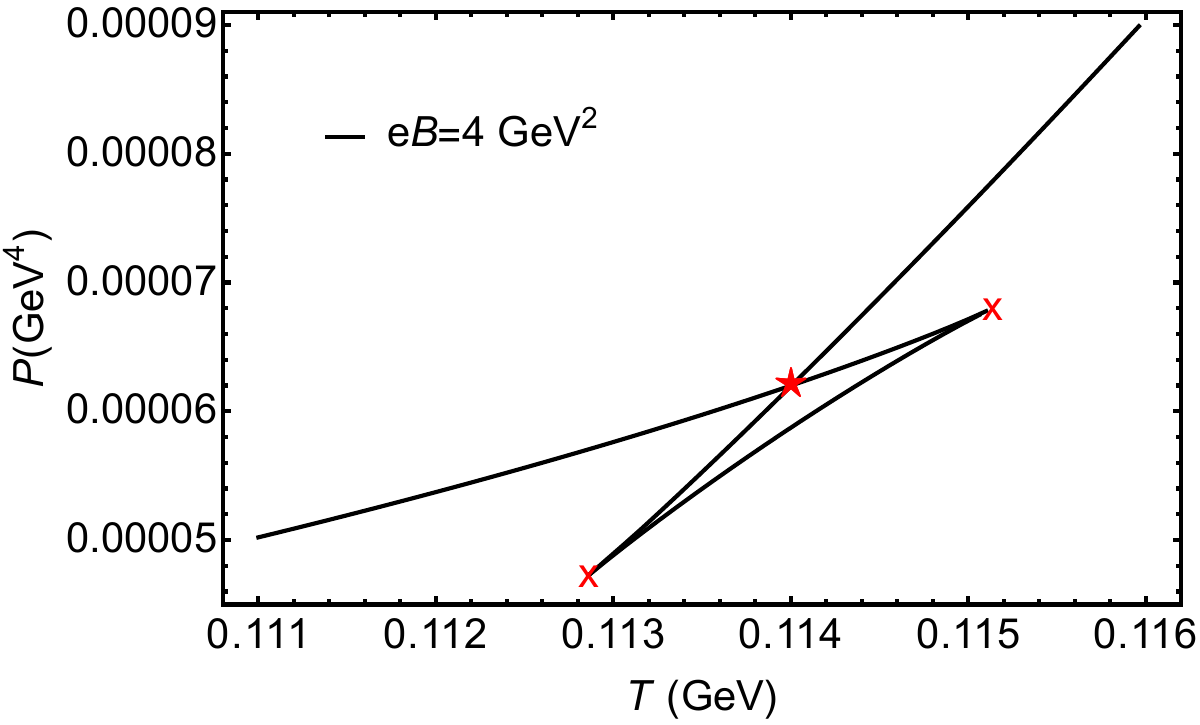}   
    \includegraphics[width=0.455\linewidth]{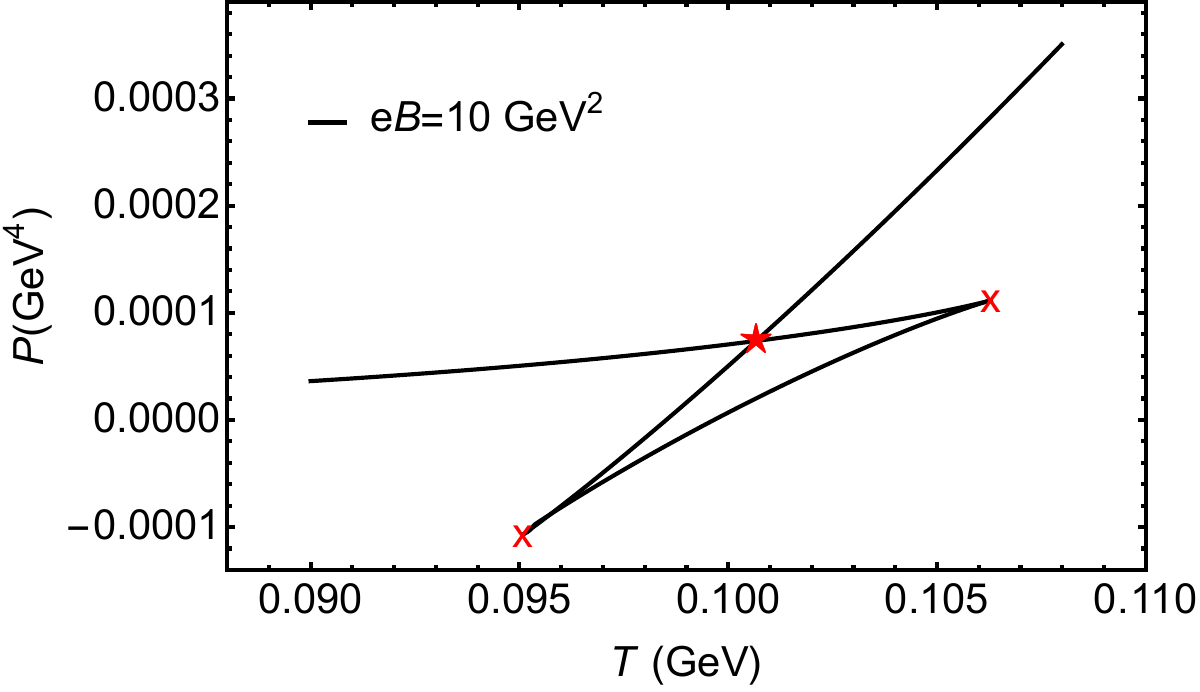}
\caption{\label{fig:pressure}  The entropy density $s$ and the pressure $P$ as functions of temperature $T$. The panels correspond to the fixed magnetic fields $eB=4 \ {\rm GeV^2}$ and the right panels  are for the fixed magnetic fields $eB=10 \ {\rm GeV^2}$.}
\end{figure*}
\section{The phase transition with finite magnetic field}
In this section, we investigate the QCD properties in a magnetic background field. The strategy is as follows: first, we  draw a vertical line to intersect with the curves in the lower panel of Fig.~\ref{fig:phibtotb}, and then numerically map from the intersection points back to determine the corresponding input parameters ($\phi_0,\mathcal{B}$). For example, as shown in Fig.~\ref{fig:mappingbackwithfixedB}, we plot three curves of the selected input parameters ($\phi_0,\mathcal{B}$), which are corresponding to the magnetic background fields set at $eB=0.1,\ 1$ and $ 10\ {\rm{GeV}^2}$, respectively. Finally, after determining several sets of the input parameters for fixed magnetic fields, we solve the EOMs in Eq.~\eqref{eq:EOMs} and obtain the corresponding values of the QCD quantities.

\subsection{The entropy and pressure}\label{sec:A}
For the mapping between ($\phi_0,\mathcal{B}$) and ($T,B$) shown in the lower panel of Fig.~\ref{fig:phibtotb}, there is a special region where a pair of ($T,B$) corresponds to different pairs of ($\phi_0,\mathcal{B}$)~\footnote{A similar result of the multiple correspondence, has been observed in Ref.~\cite{DeWolfe:2010he}, in which the authors studied the critical point at finite temperature and chemical potential.}. Specifically, as $eB$ increases from small to large, this special region begins at a particular point. This behavior could be a signal of the first-order phase transition. To confirm this conjecture, we  study the behaviors of the entropy density $s$ and pressure $P$ in this region in the following part.

Entropy is an extensive quantity, which can be extracted through the Bekenstein-Hawking relation as $S=A/4G_N$ with $A$ the area of the horizon. Thus, the entropy density can be obtained by dividing the volume $V$,
\begin{equation}
    {s}=\frac{S}{V}=\frac{2\pi}{\kappa^2}e^{\tilde{a}(\tilde{r}_H)+2\tilde{c}(\tilde{r}_H)}=\frac{2\pi e^{2(a_0^{\rm far}-c_0^{\rm far})}}{\kappa^2\phi_A^{3/\nu}}.
\end{equation}

In  the upper panel of Fig.~\ref{fig:entropy}, we show the numerical results of the temperature dependence of the normalized entropy density $s/T^3$ for four different fixed magnetic fields, $eB=0,\ 1,\ 2.8326,$ and $ 4$ $\rm GeV^2$.  In the absence of a magnetic background field, the normalized entropy density is small at low temperatures and exhibits a monotonic increase as the temperature rises.  At high temperatures, it behaves according to the relation $s\sim T^3$. This may indicate that the QCD system undergoes a crossover from the low-temperature normal phase to the high-temperature quark-gluon-plasma (QGP) phase. 
With a finite magnetic background field,  the normalized entropy density in both the low and high-temperature limits consistently approaches the values observed in the absence of a magnetic field. In the intermediate temperature region, there is a noticeable enhancement of the normalized entropy density, which becomes more pronounced with an increase in the magnetic field. When $eB\gtrsim 2.8326\ {\rm GeV^2}$, as shown in the inset in the upper panel of Fig.~\ref{fig:entropy}, the normalized entropy density as a function of temperature exists multiple value region, signaling the onset of the first-order phase transition. When $eB\leq 2.8326\ {\rm GeV^2}$, we shown the specific heat $C=T\partial s/\partial T$ with $\mu_B=0$ in the lower panel of Fig.~\ref{fig:entropy}. It diverges at a certain temperature $T_c\approx 0.1191 \ {\rm GeV}$ with the critical magnetic field $eB_c=2.8326\ {\rm GeV^2}$. Below  $eB_c=2.8326\ {\rm GeV^2}$, the susceptibility of entropy density with respect to the temperature is smooth and continuous, indicating a crossover transition. Therefore, the point $(eB_c, T_c)=(2.8326\ {\rm GeV^2}, 0.1191 \ {\rm GeV})$ is a critical endpoint, corresponding to a second-order phase transition, where the first-order transition changes to a crossover.

\begin{figure}[tbh]
    \centering
    \includegraphics[width=0.9\linewidth]{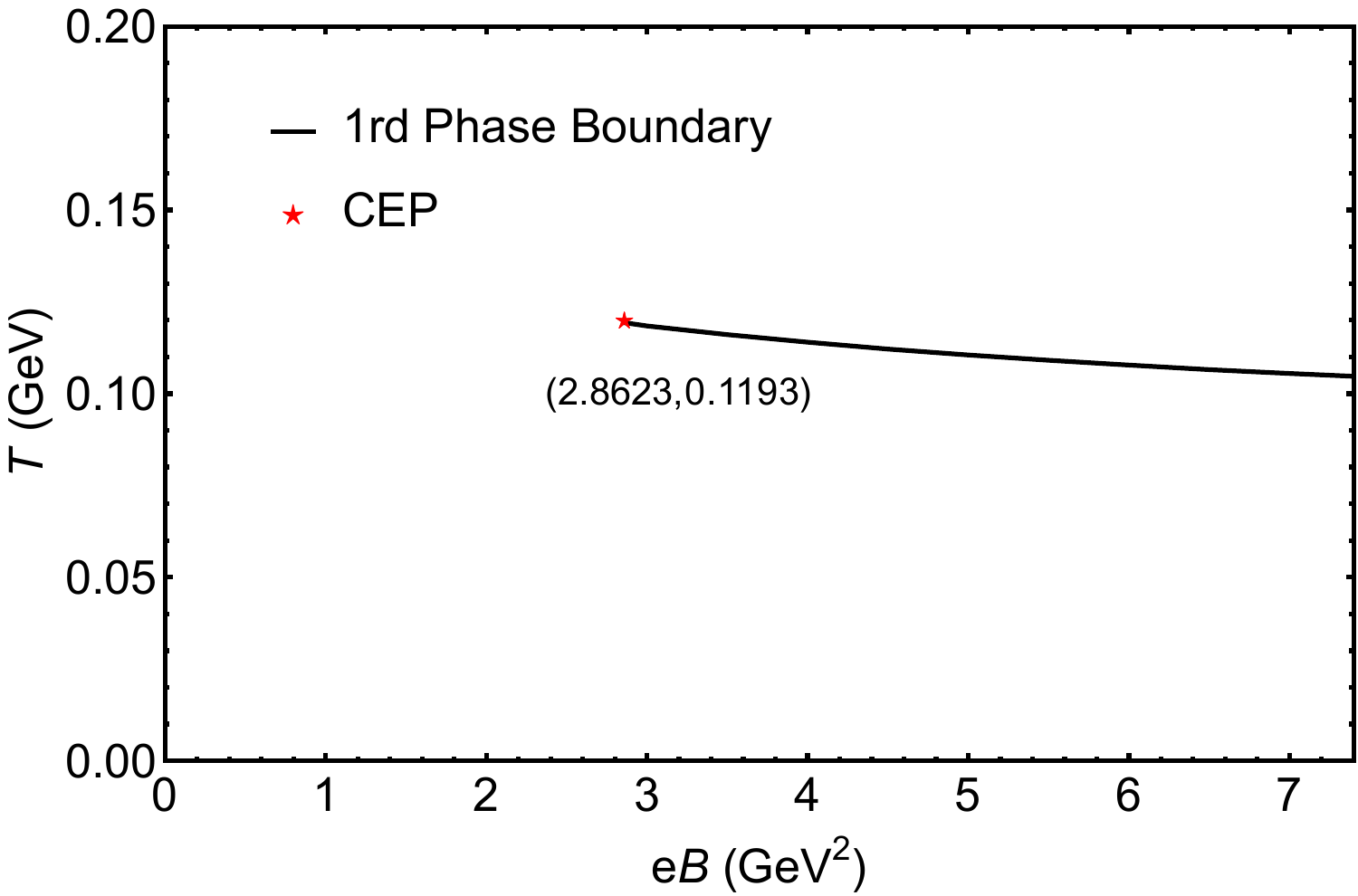}
\caption{The phase diagram in the plane of magnetic field and temperature.The critical end point (CEP) of the first order phase boundary is located at $(eB_c,T_c)=(2.8623\ {\rm GeV^2}, 0.1193\ {\rm GeV})$}\label{fig:phasediagram}
\end{figure}

To determine the phase boundary of the first-order transition, we study the behavior of the pressure $P$. At a fixed magnetic background field, the pressure can be approximately determined as
\begin{eqnarray}
P(T)\approx \int_{T_{\rm low}}^{T} s(T)  dT,
\end{eqnarray}
where we select a low temperature $T_{\rm low}$ as the lower boundary rather than absolute zero due to the impracticality of achieving zero temperature within this model. This selection introduces only a constant difference and does not affect the results of our analysis. For convenience, we choose $T_{\rm low}=0.053\ {\rm GeV}$ in our calculations. In Fig.~\ref{fig:pressure}, we show the temperature dependence of the entropy density (S-shape) and pressure (Swallowtail shape) in the upper panels and lower panels, respectively. The corresponding turning points of the entropy and pressure are labeled with red cross symbols. The first-order phase transition temperature and the unstable region can be  determined  by employing the Maxwell construction, where the areas separated by the isotherm in the thermodynamic curves are equal.  In the lower left panel, the five-pointed star symbol determines the transition temperature $T_t=0.114\ {\rm GeV}$,  at which the low-temperature branch of the entropy density curve directly jumps to the high-temperature branch through the red vertical line of left upper panel, with the remaining part representing the unstable or nonphysical states. For any other magnetic background field $eB>2.8326\ {\rm GeV^2}$, we can also determine the transition temperature using this method. Finally, we show the phase diagram in the magnetic field-temperature plane in Fig.~\ref{fig:phasediagram},  where the five-points star symbol represents the critical endpoint (CEP) and the black line represents the first-order boundary.  As the results demonstrate, the temperature at which the first-order phase transition occurs diminishes with an increase of the magnetic field.

\subsection{Polyakov loop}
In the previous subsection, we analyzed the phase transition from the perspective of entropy and pressure. However, to explicitly determine the type and symmetry breaking associated with the phase transition, it is essential to consider the Polyakov loop which is the order parameter of the confienment/deconfinement phase transition. The Polyakov loop, as derived in the holographic framework, was first extracted by Noronha in Ref.~\cite{Noronha:2009ud}. The expectation value of the Polyakov loop operator is defined as follows:
\begin{eqnarray}
    P\equiv |\langle\hat{L}_P\rangle|=e^{-F_Q/T}
\end{eqnarray}
with $F_Q$ the heavy quark free energy~\cite{Noronha:2010hb}. The heavy quark free energy is given by~\cite{Noronha:2009ud,Critelli:2017oub,Rougemont:2023gfz},
\begin{eqnarray}\label{eq:polyakovloop}
    F_Q &=& \frac{\sqrt{\lambda}}{2\pi}\left [\int_{\tilde{r}_H}^{\infty}dr \sqrt{-\tilde{g}_{tt}^{(s)}\tilde{g}_{rr}^{(s)}}-\sqrt{\mathrm{Asy}\{-\tilde{g}_{tt}^{(s)}\tilde{g}_{rr}^{(s)}\}}\right.\nonumber\\
    & &\left.-\int_{\rm cte}^{\tilde{r}_H} dr  \sqrt{\mathrm{Asy}\{-\tilde{g}_{tt}^{(s)}\tilde{g}_{rr}^{(s)}\}}\right ],
\end{eqnarray}
where the upper indicator $(s)$ for the string frame with $\tilde{g}_{\mu\nu}^{(s)}=e^{\sqrt{2/3}\phi}\tilde{g}_{\mu\nu}$; \textrm{Asy}$\{-\tilde{g}_{tt}^{(s)}\tilde{g}_{rr}^{(s)}\}$ is the asymptotic value for large $r$, which equals to $e^{2r}$; $r_{\rm cte}$ is dependent on the renormalization scheme. Substituting coordinate transformation Eq.~\eqref{eq:explicitrelation} into Eq.~\eqref{eq:polyakovloop}, one has~\footnote{In the numerical process, the integration interval is chosen to be $[\epsilon,r_{max}]$.}
\begin{eqnarray}
    F_Q &=& \frac{\sqrt{\lambda}}{2\pi}\left [\int_{{r}_H}^{r_{max}}dr \frac{(e^{\sqrt{2/3}\phi(r)+a(r)}-e^{r/h_0^{\rm far}+a_0^{\rm far}})}{\phi_A^{1/\nu}}\right.\nonumber\\
    & &\left.-\int_{r_{\rm cte}}^{{r}_H} dr  \frac{e^{r/h_0^{\rm far}+a_0^{\rm far}}}{\phi_A^{1/\nu}}\right ],
\end{eqnarray}
We adopt the renormalization scheme in Ref.~\cite{Bruckmann:2013oba} as
\begin{eqnarray}
    F_Q^r=F_Q(T,B)-F_Q(T_{\rm x},0),
\end{eqnarray}
where $T_{\rm x}$ is the fixed temperature. We take the fixed temperature $T_{\rm x}=360$ MeV. Thus, renormalization Polyakov loop,
\begin{eqnarray}
    P_r=e^{-F_Q^r/T}.
\end{eqnarray}

\begin{figure*}
    \includegraphics[width=0.43\textwidth]{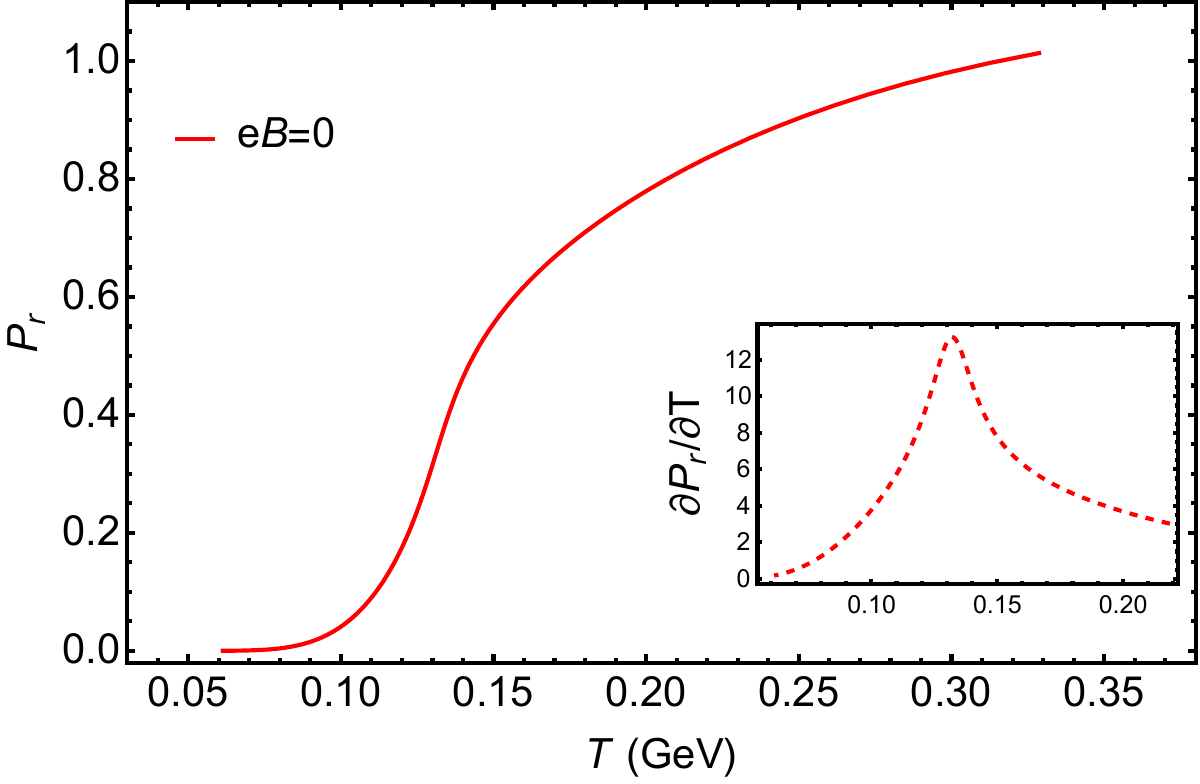}
    \hspace{0.5cm}
    \includegraphics[width=0.43\textwidth]{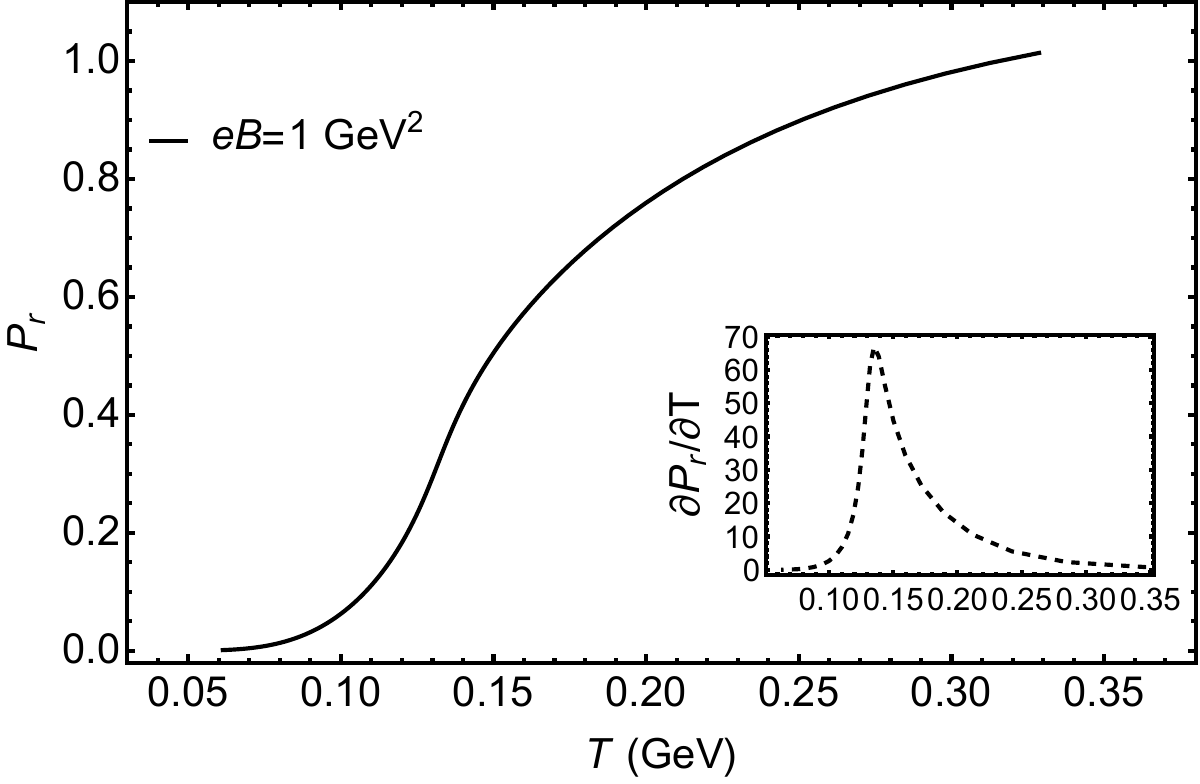}
    \includegraphics[width=0.43\textwidth]{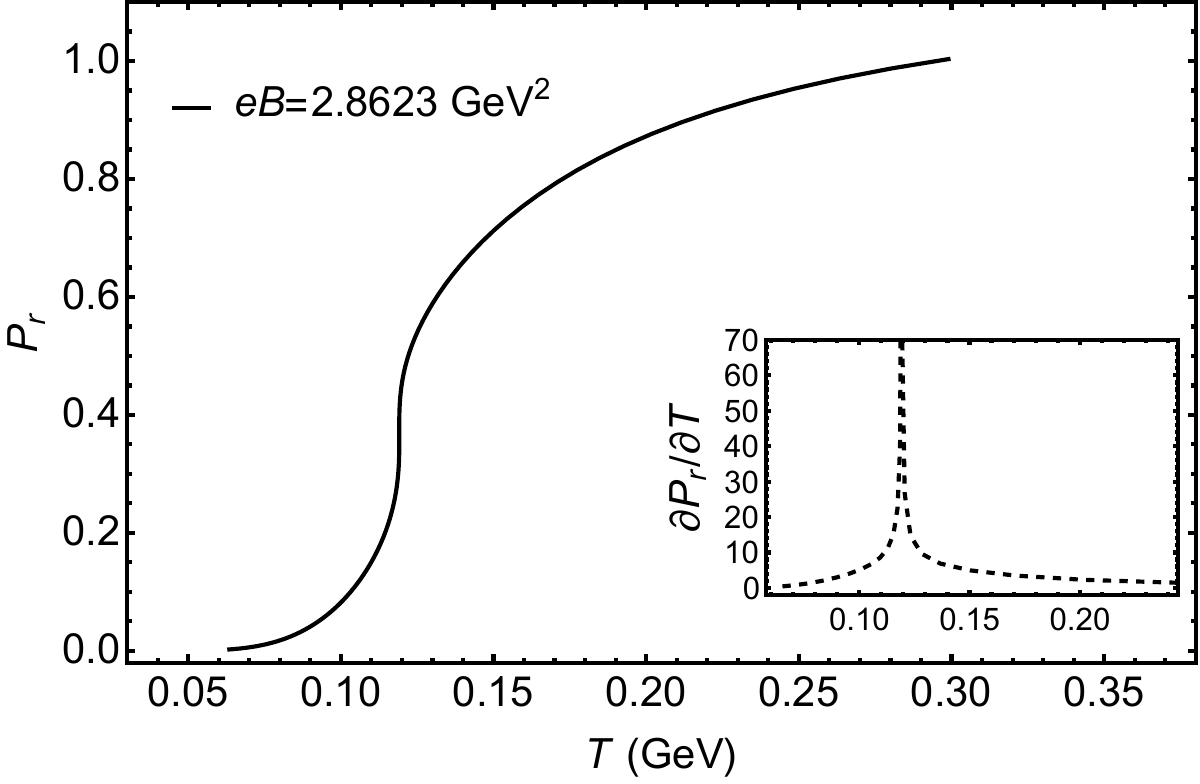}
    \hspace{0.5cm}
    \includegraphics[width=0.43\textwidth]{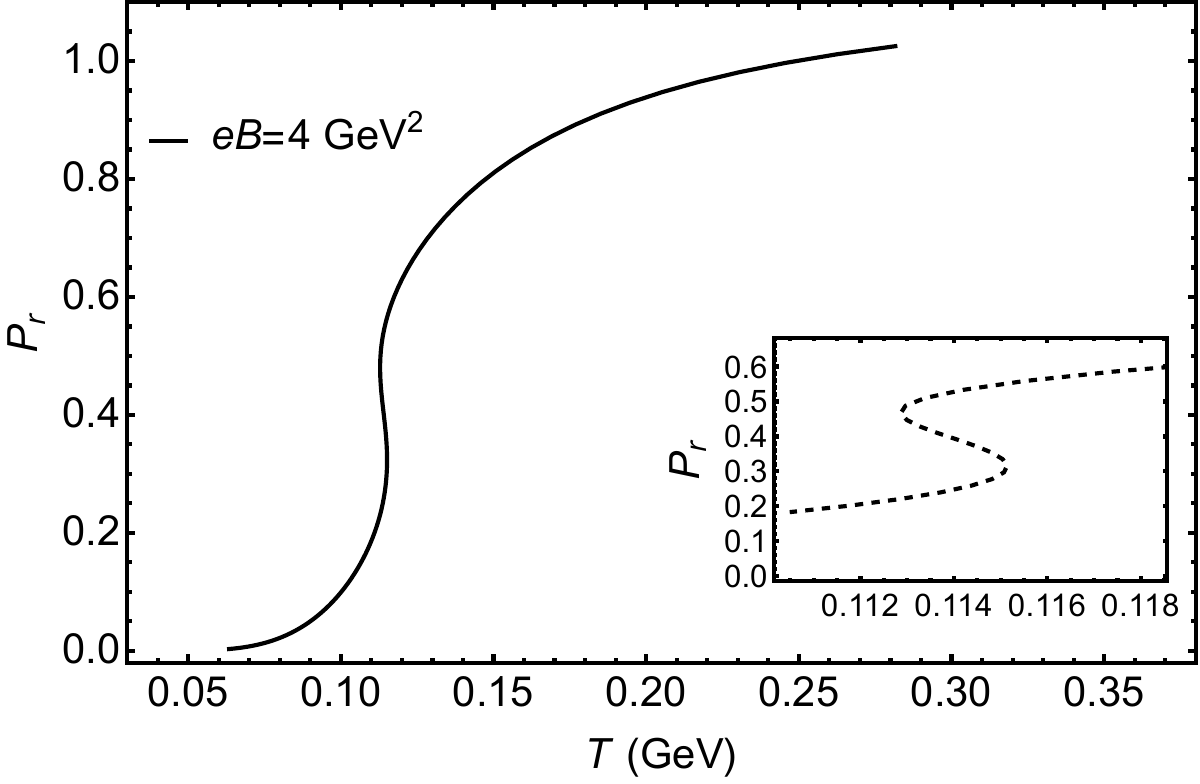}
    \caption[short]{ The renormalized Polyakov loop, $P_r$, is shown as a function of temperature for different magnetic field strengths: $B = 0$, $eB = 1 , \rm{GeV}^2$, $eB = 2.8623 , \rm{GeV}^2$, and $eB = 4 , \rm{GeV}^2$, arranged from left to right and top to bottom. The inset figures in the first three subfigures display $\partial P_r/\partial T$ in the region where $P_r$ rises rapidly with increasing temperature. For the last subfigure, the inset provides a zoomed-in view of the same region to highlight detailed behavior.}\label{fig:polyakov_loop}
\end{figure*}
To elucidate the change in the order of the confinement/deconfinement phase transition induced by the magnetic field, we calculate the Polyakov loop at four specified magnetic field strengths: $B = 0$, $1$, $2.8623$, and $4$ $\rm GeV^2$. The results for the renormalized Polyakov loop, $P_r$, and its first derivative with temperature, $\partial P_r/\partial T$, are presented in Figure \ref{fig:polyakov_loop}.

At zero magnetic field ($eB = 0$), $P_r$ transitions smoothly from a small value near zero to a larger value as the temperature increases. The first derivative of $P_r$ with respect to temperature, $\partial P_r/\partial T$, rises to a peak and then decreases, as shown in the inset of the figure. This behavior indicates a crossover from the confined phase to the deconfined phase. At $eB = 1 \ \rm GeV^2$, both $P_r$ and its temperature derivative exhibit similar behavior to the case at $eB = 0$.
For $eB = 2.8623 \ \rm GeV^2$, the Polyakov loop exhibits a second-order phase transition. In this case, $\partial P_r/\partial T$ diverges at $T = 0.119$ GeV, which corresponds to the critical endpoint (CEP) identified from the entropy density and pressure analysis in Sec.\ref{sec:A}. Comparing $P_r$ at the same temperature for different magnetic field strengths, we observe that a stronger magnetic field leads to a larger $P_r$, consistent with the lattice QCD results reported in Ref.\cite{Bruckmann:2013oba}.
As shown in the last subfigure of Figure \ref{fig:polyakov_loop}, the behavior of $P_r$ at $eB = 4 \ \rm GeV^2$ differs significantly from the magnetic field below the CEP ($eB \leq 2.8623 \ \rm GeV^2$). In this regime, $P_r$ exhibits an S-shaped curve with multiple values around $T = 0.114$ GeV, indicating a first-order phase transition for the confinement/deconfinement transition. This observation is consistent with the conclusions derived from the entropy density and pressure in Sec.~\ref{sec:A}.
Thus, the presence of a magnetic field significantly impacts the confining properties of QCD, driving the transition from a crossover to a first-order phase transition as the magnetic field strength increases.

\section{Jet quenching parameter}
From the study of the phase transition in the temperature and magnetic field plane in the frame of holographic QCD, we find there are crossover, first-order and second-order phase transitions. It would be interesting to investigate the jet quenching parameter within these different phase transition backgrounds. This investigation of jet quenching parameter is essential for gaining insights into the QGP from both theoretical and experimental perspectives, as well as for understanding the phase transitions within the framework of QCD.  Although laboratory-generated magnetic fields may not reach extremely high values, the investigation for the extremely magnetic field region remains meaningful. It is important to note that the magnetic field generated in heavy-ion collisions is not constant but highly time-dependent. The results presented in this paper are based on the assumption of a constant magnetic field, which provides an idealized estimate of the QCD phase structure and jet quenching behavior. While this simplification facilitates theoretical calculations, extending the model to account for the transient nature of the magnetic field remains a crucial task for bridging the gap between theoretical predictions and experimental observations. On one hand, relevant phenomena can be compared to the high density case. On the other hand, extremely strong magnetic fields might have existed in the early universe making the study of first-order transition very important.

\begin{figure}[tbh]
    \centering    \includegraphics[width=0.9\linewidth]{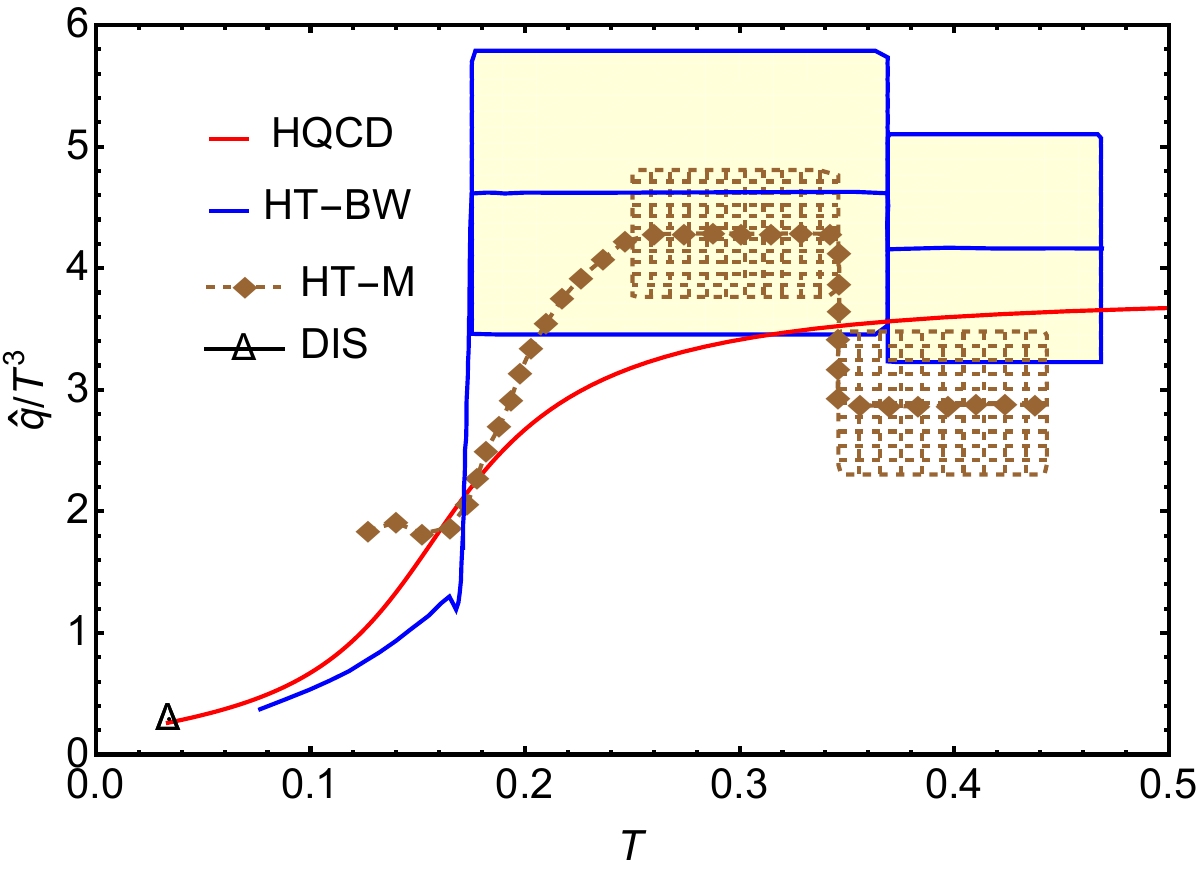}
\caption{The normalized jet quenching parameter $\hat{q}/T^3$ as a function of temperature without magnetic background field $eB=0$. The red curve is the holographic QCD result with $\lambda_t=1/2$. The HT-BW, HT-M model results, for an initial quark jet with energy $E=10$ GeV, and deep inelastic scattering (DIS) results are all taken from  Refs.~\cite{JET:2013cls,Cao:2020wlm}. The filled boxes and meshes represent for the errors.}\label{fig:jqb0}
\end{figure}

\begin{figure*}[tbh]
\centering
\includegraphics[width=0.3\linewidth]{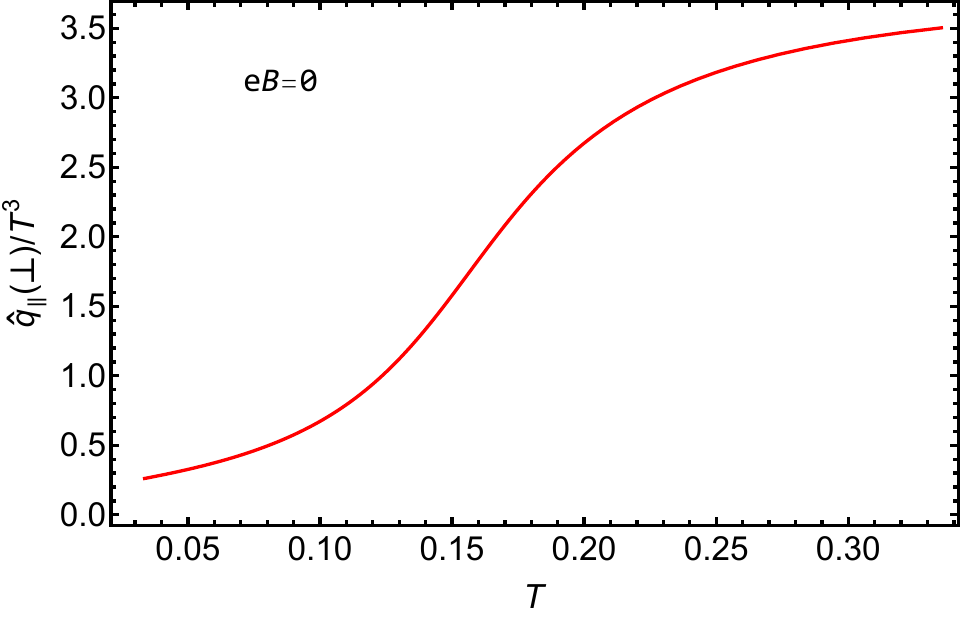}
\includegraphics[width=0.3\linewidth]{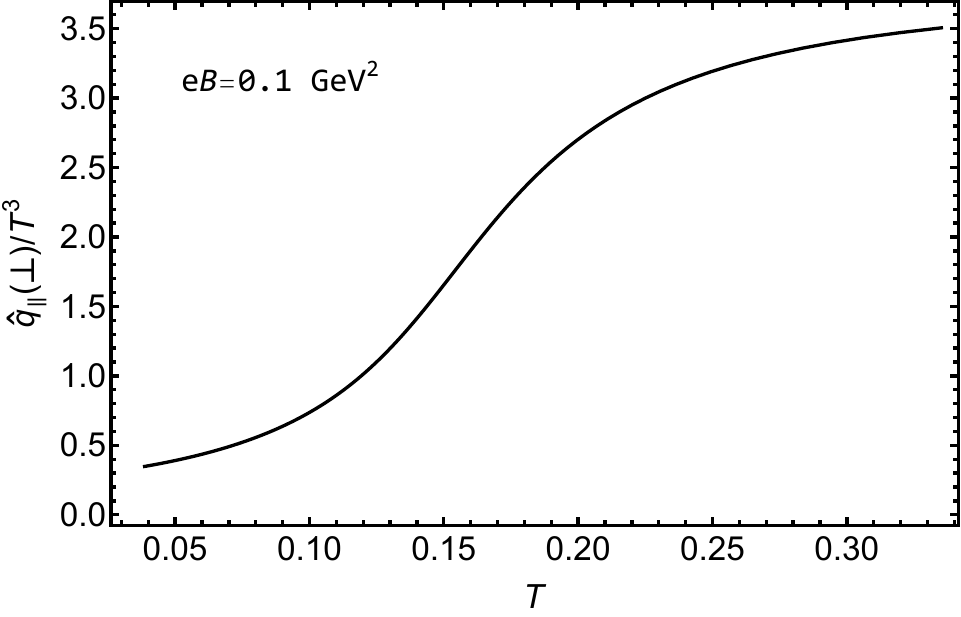}
\includegraphics[width=0.3\linewidth]{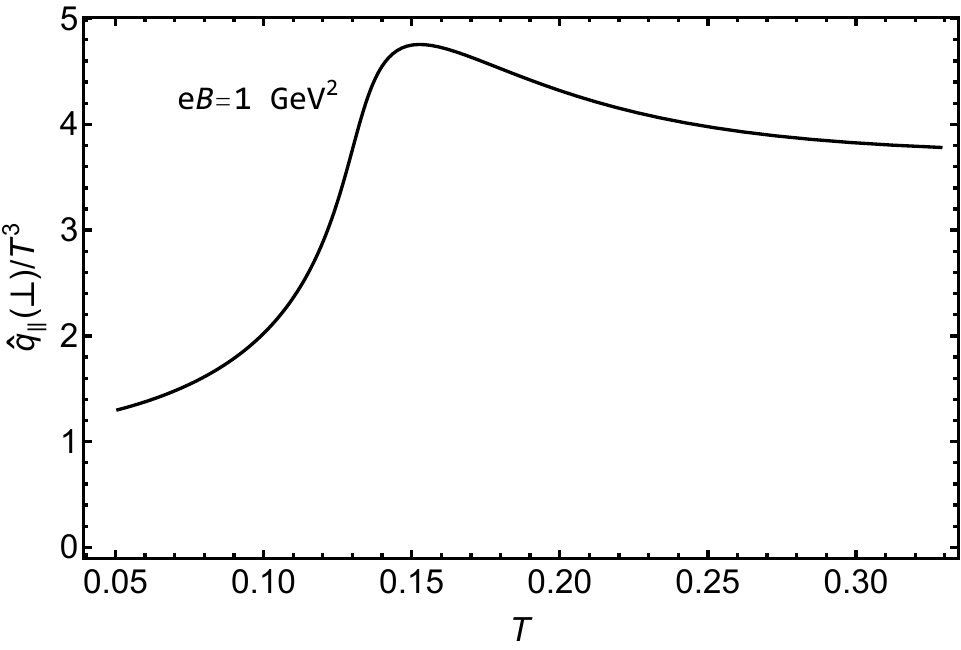}
\includegraphics[width=0.3\linewidth]{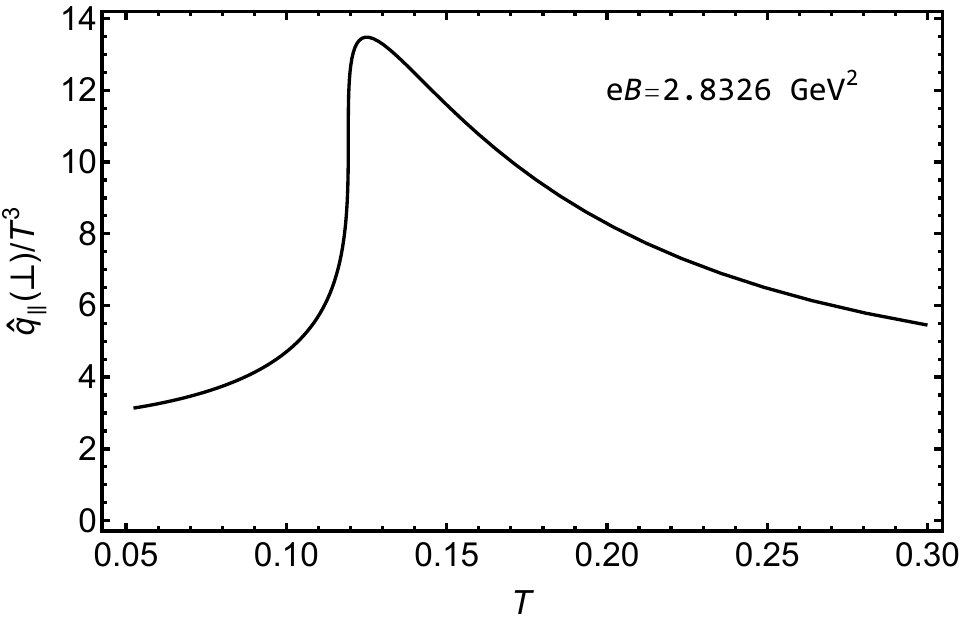}
\includegraphics[width=0.3\linewidth]{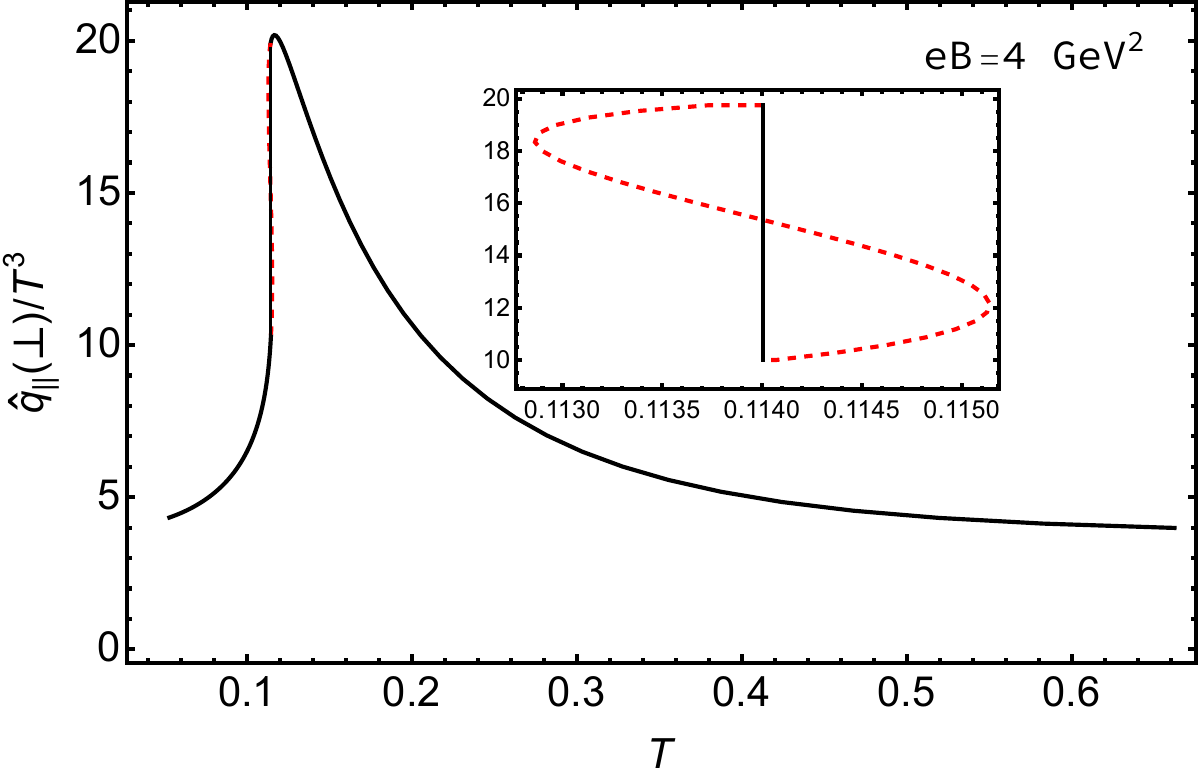}
\includegraphics[width=0.3\linewidth]{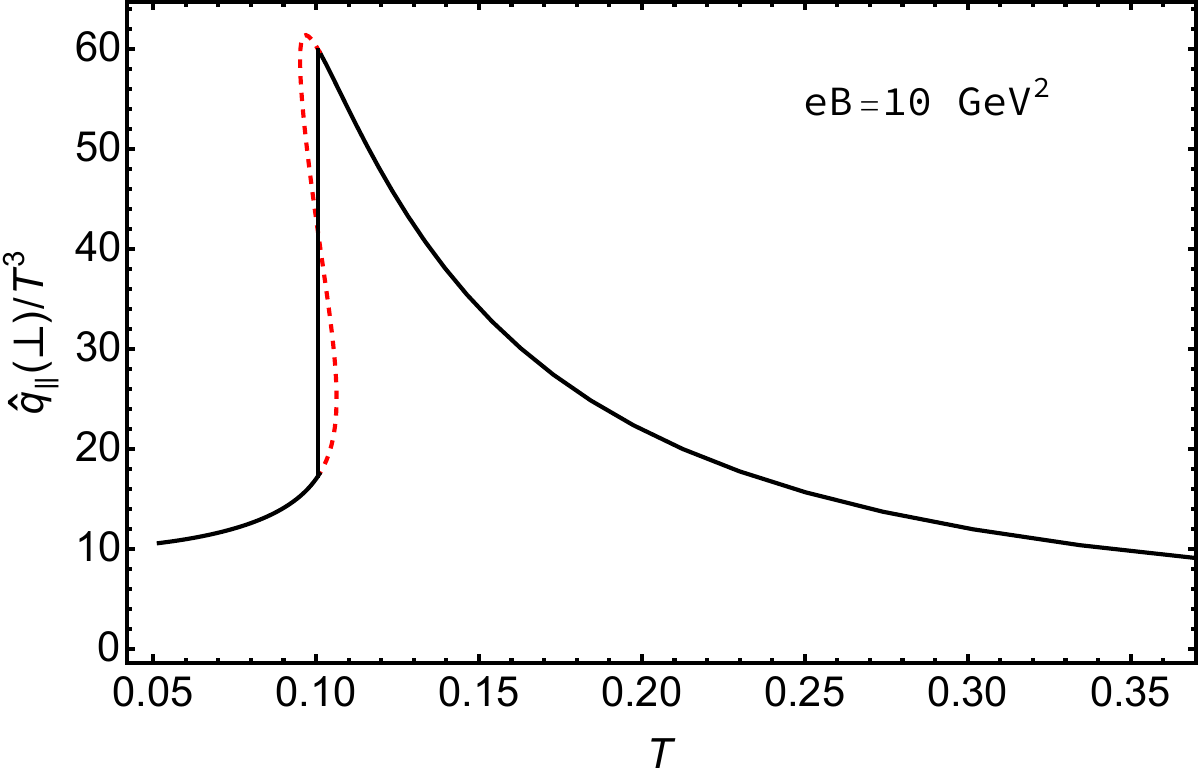}
\caption{\label{fig:jqzx} With different fixed magnetic fields, the temperature dependence of the jet quenching parameters $\hat{q}_{||}(\perp)$ for the parton moving in the direction of the magnetic field and the transverse momentum broadening in the perpendicular to the magnetic field.}
\end{figure*}

\begin{figure*}[htb]
\centering
\ \ \includegraphics[width=0.45\linewidth]{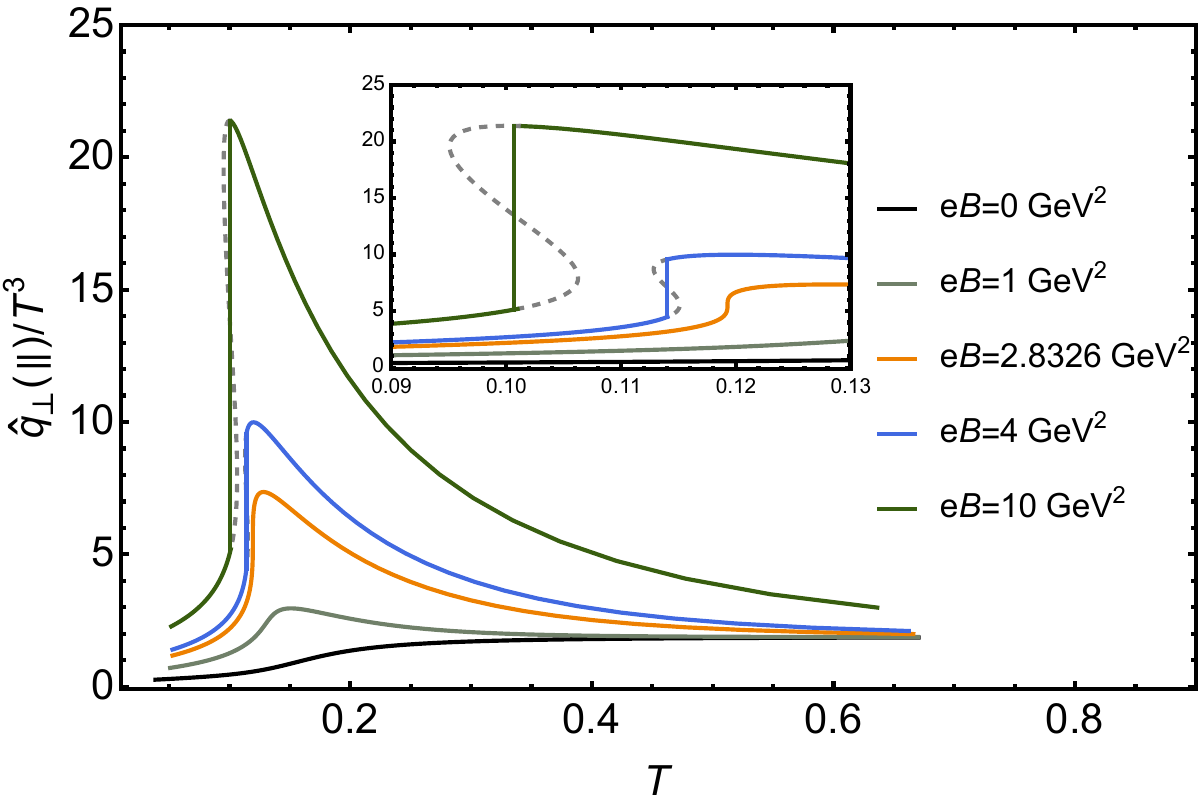}
\includegraphics[width=0.46\linewidth]{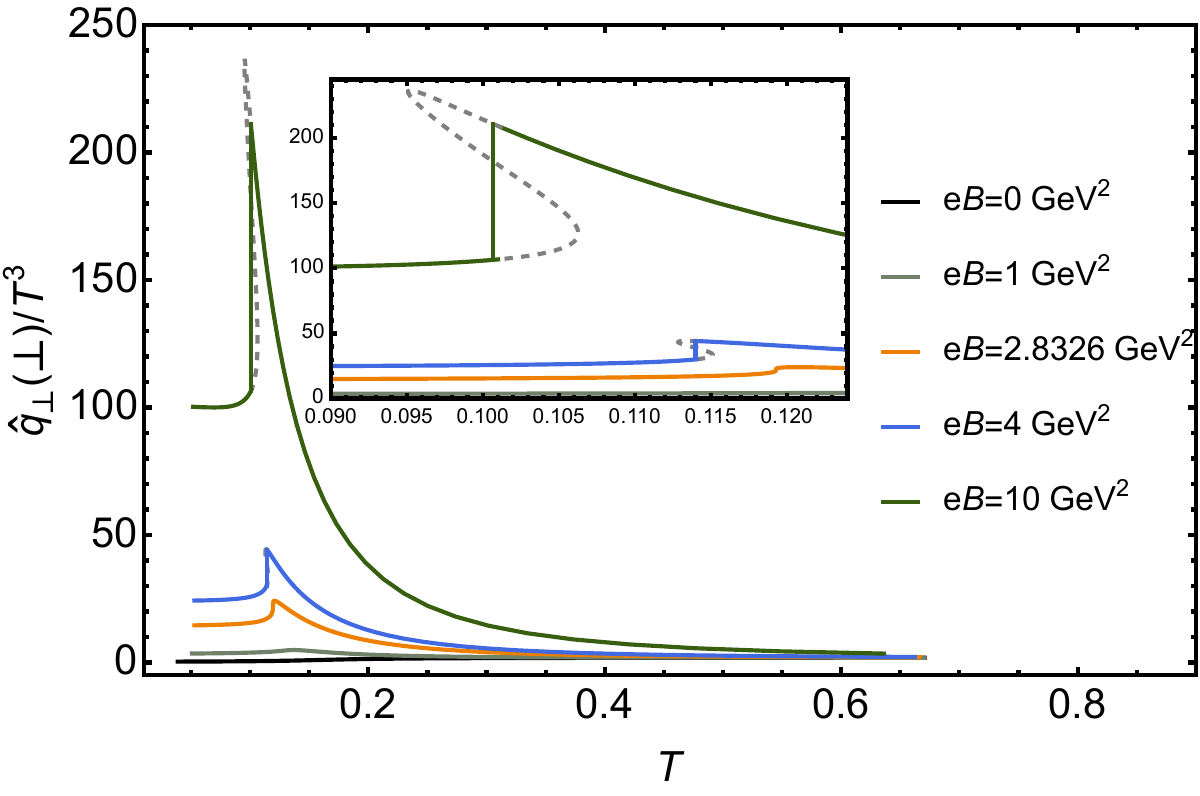}
\caption{\label{fig:jqxz} For the left panel, the jet quenching parameter $\hat{q}_{\perp}(||)$ as a function of temperature under different fixed magnetic field. For the right panel, the jet quenching parameter $\hat{q}_{\perp}(\perp)$ as a function of temperature under different fixed magnetic field. The inserts are the enlarge of the transition temperature regions.}
\end{figure*}

Within the framework of holographic QCD, the calculation of the jet quenching parameter was first obtained by Hong Liu, Krishna Rajagopal, and Urs Achim Wiedemann in Ref.~\cite{Liu:2006ug}. Since then, many studies have been conducted under various conditions and using different holographic models. However, the jet quenching parameter in extremely strong magnetic fields remains unexplored. Our calculations suggest that the phase transition in this condition could be of the first-order,  thereby rendering the discussion of jet quenching parameters in this scenario both compelling and significant.

Due to the magnetic background field, the spatial $SO(3)$ symmetry is broken down to $SO(2)$. In the anisotropic medium, there are three different types of jet quenching parameters. First, the light parton moves along the direction of the magnetic field, with transverse momentum broadening perpendicular to the magnetic field direction. Second, the light parton moves perpendicular to the magnetic field, with transverse momentum broadening in the magnetic field direction. Third, the light parton moves perpendicular to the magnetic field direction, with transverse momentum broadening also perpendicular to the magnetic field direction.
Following the steps in Refs.~\cite{Liu:2006ug,Rougemont:2020had}, one can derive the jet quenching parameters $\hat{q}$ as
\begin{eqnarray}\label{eq:jetquenchingp}
    \hat{q}_p(k)=\frac{\sqrt{\lambda_t}}{\pi}\left(\int_{\tilde{r}_H}^\infty d\tilde{r} \frac{1}{\tilde{g}_{kk}}\sqrt{\frac{\tilde{g}_{rr}}{\tilde{g}_{tt}+\tilde{g}_{pp}}}\right)^{-1},
\end{eqnarray}
where $p$ is the direction of the light parton moving and $k$ is the direction of the transverse momentum broadening. $\lambda_t$ is the 't Hooft coupling. Substituting the relation Eq.~\eqref{eq:explicitrelation} into the expression of the jet quenching parameter Eq.~\eqref{eq:jetquenchingp}, one can obtain the expressions for these normalized jet quenching parameters.
\begin{widetext}
\begin{subequations}
    \begin{eqnarray}
       \frac{\hat{q}_{||}(\perp)}{T^3}= \frac{64\pi^2 \sqrt{\lambda_t} h_0^{\rm far}}{\int_{r_H}^{r_{max}}d r\frac{e^{-2c(r)-a(r)+2(c_0^{\rm far}-a_0^{\rm far})}}{\sqrt{h(r)[h_0^{\rm far}-h(r)]}}}\\
        \frac{\hat{q}_{\perp}(||)}{T^3}=\frac{32\pi^2 \sqrt{\lambda_t} h_0^{\rm far}}{\int_{r_H}^{r_{max}}d r\frac{e^{-2a(r)}}{\sqrt{h(r)[h_0^{\rm far}e^{2c(r)-2(c_0^{\rm far}-a_0^{\rm far})}-h(r)e^{2a(r)}]}}}\\
        \frac{\hat{q}_{\perp}(\perp)}{T^3}=\frac{32\pi^2 \sqrt{\lambda_t} h_0^{\rm far}}{\int_{r_H}^{r_{max}}d r\frac{e^{-2c(r)+2(c_0^{\rm far}-a_0^{\rm far})}}{\sqrt{h(r)[h_0^{\rm far}e^{2c(r)-2(c_0^{\rm far}-a_0^{\rm far})}-h(r)e^{2a(r)}]}}}
    \end{eqnarray}
\end{subequations}
\end{widetext}

In Fig.~\ref{fig:jqb0}, we show the results of $\hat{q}$ without a magnetic field.
As shown by the red curve, we have the normalized jet quenching parameter $\hat{q}/T^3$ as a function of the temperature. In our calculations, we choose the 't Hooft coupling $\lambda_t=1/2$.  We have found that the holographic outcomes are in agreement with the Monte Carlo simulations, which were conducted using an initial quark jet with an energy of  $E=10$ GeV, as well as the results from the deep inelastic scattering (DIS) ~\cite{JET:2013cls,Cao:2020wlm}. The holographic curve lies within the region of the Monte Carlo and the DIS results. The significant difference between the low and high temperatures indicates a crossover from the normal phase to the QGP phase.

In Fig.~\ref{fig:jqzx}, we show the normalized jet quenching parameter $\hat{q}_{||}(\perp)$ as a function of temperature for different fixed magnetic background fields. Comparing the sub-figures, one can see that $\hat{q}_{||}(\perp)$ is generally enhanced by the magnetic field, favoring  the transverse momentum broadening. This enhancement is especially significant in the phase transition temperature region and tends to match the values of the zero magnetic field case in both the low and high-temperature limits. Furthermore, with relatively small magnetic fields, $\hat{q}_{||}(\perp)/T^3$ remains a monotonic function of temperature.  However,  in the presence of an extremely strong magnetic field, for instance, $eB=4$ and $\ 10\ {\rm GeV^2}$, $\hat{q}_{||}(\perp)/T^3$, there exists a region around the phase transition temperature where $\hat{q}_{||}(\perp)/T^3$ takes on multiple values. This results in three different values of $\hat{q}_{||}(\perp)/T^3$ for a single temperature, as indicated by the red dashed lines in Fig.~\ref{fig:jqzx}, which is the typical characteristic of the first-order transition. At the critical value of the magnetic field, i.e., $eB=2.8326\ {\rm GeV^2}$, the first-order derivative of jet quenching parameter with respect to the temperature $\partial \hat{q}_{||}(\perp)/\partial T$ diverges at $T_c$.

For the other two types of jet quenching parameters, $\hat{q}_{\perp}(||)$ and $\hat{q}_\perp(\perp)$, the numerical results are shown in Fig.~\ref{fig:jqxz}. The overall trends of these jet quenching parameters are almost the same as $
\hat{q}_{||}(\perp)$ shown in Fig.~\ref{fig:jqzx}. Specifically, $\hat{q}_{\perp}(||)+\hat{q}_{\perp}(\perp)$ is larger than $\hat{q}_{||}({\perp})$; $\hat{q}_{\perp}(\perp)$ is larger than $\hat{q}_\perp({||})$ under the same temperature and magnetic field conditions~\footnote{This is consistent with the results in a narrower region of magnetic field  \cite{Rougemont:2020had}}.


\section{CONCLUSIONS}
In this paper, we investigate the impact of a magnetic background field on QCD properties using the EMD model. Our findings reveal that at extremely high magnetic fields, the QCD phase transition is of first-order. As the magnetic field decreases, the first-order phase transition boundary terminates at a critical endpoint, located at $(eB_c,T_c)= (2.8326\ {\rm GeV^2}, 0.1191 \ {\rm GeV})$, before which the transition turns into a crossover.  The entropy density and pressure behavior confirmed this phase transition boundary, with typical S-shape and Swallowtail shape curves. The first-order transition temperature decreases with the increasing of magnetic field. These are qualitatively consistent with lattice results in Ref.~\cite{DElia:2021yvk}. Additionally, both the entropy and pressure are significantly enhanced by the magnetic fields.  The analysis of the Polyakov loop confirms that the confinement/deconfinement phase transition evolves with increasing magnetic field strength, transitioning from a crossover to a first-order phase transition. This observation, consistent with entropy density and pressure results.

Furthermore, we also discuss the variation of the jet-quenching parameter during the phase transition process in the presence of a magnetic field. For comparison, in the absence of a magnetic field, our holographic results are primarily consistent with the Monte Carlo and the DIS experiment ~\cite{JET:2013cls,Cao:2020wlm}, provided that $\lambda_t=1/2$. In the presence of a magnetic field, it highlights the anisotropic nature of QGP  in heavy ion collisions. The magnetic field reduces the spatial $SO(3)$ symmetry to $SO(2)$, leading to distinct jet quenching parameters that vary with the direction of the parton’s motion and the orientation of the transverse momentum broadening. Under the phase transition, the jet quenching parameters are greatly affected by the magnetic field. The normalized jet quenching parameters are enhanced around the phase transition temperature, suggesting that magnetic fields promote the transverse momentum broadening as well as the jet energy loss in the QGP. The parameters converge to the zero magnetic field cases at both low and high-temperature extremes. Moreover, as a consequence of the first-order phase transition, they can exhibit multiple values which indicates that jet quenching could serve as a signal to probe the phase diagram of QCD phase transition.

\section*{Acknowledgement}
We thank Danning Li for his beneficial discussion. This work is supported by the Guangdong Basic and Applied Basic Research Foundation under Grant No. 2024A1515012931, the National Natural Science Foundation of China (NSFC) under Grant Nos. 12305142, 12275108  and the Science and Technology Planning Project of Guangzhou, China under Grant No. 2024A04J3243.
\bibliographystyle{apsrev4-1}
\bibliography{magnetbib}

\end{document}